\documentclass[aps,prb,twocolumn,superscriptaddress,floatfix,amsmath,amssymb]{revtex4-2}
\usepackage{amsmath}
\usepackage{amssymb}
\usepackage{braket}
\usepackage{graphicx}
\usepackage[table]{xcolor}
\usepackage{verbatim}
\usepackage{float}
\usepackage{color}
\usepackage{siunitx}
\usepackage{gensymb}
\usepackage{bm}
\usepackage{xcolor}
\usepackage{multirow}
\usepackage{footnote}
\usepackage{bbm}

\usepackage{array}
\newcolumntype{P}[1]{>{\centering\arraybackslash}p{#1}}
\newcolumntype{Q}[1]{>{\raggedleft\arraybackslash}p{#1}}
\newcolumntype{R}[1]{>{\raggedright\arraybackslash}p{#1}}

\newcommand{\rr}{\mathbf{r}}
\newcommand{\RR}{\mathbf{R}}
\newcommand{\qq}{\mathbf{q}}
\newcommand{\kk}{\mathbf{k}}
\newcommand{\vv}{\mathbf{v}}

\newcommand{\GG}{\mathbf{G}}
\newcommand{\gG}{\mathbf{g}}
\newcommand{\uvec}[1]{\hat{\mathbf{\boldsymbol{#1}}}}

\newcommand{\braketD}[2]{\langle #1 | #2 \rangle}

\begin{document}
\title{Moir\'e band structures of twisted phosphorene bilayers}

\author{Isaac Soltero}
\affiliation{Departamento de F\'isica Qu\'imica, Instituto de F\'isica, Universidad Nacional Aut\'onoma de M\'exico, Ciudad de M\'exico, C.P. 04510, M\'exico}

\author{Jonathan Guerrero-S\'anchez}
\affiliation{Centro de Nanociencias y Nanotecnolog\'ia, Universidad Nacional Aut\'onoma de M\'exico, Apdo.\ Postal 14, 22800 Ensenada, Baja California, M\'exico}

\author{Francisco Mireles}
\affiliation{Centro de Nanociencias y Nanotecnolog\'ia, Universidad Nacional Aut\'onoma de M\'exico, Apdo.\ Postal 14, 22800 Ensenada, Baja California, M\'exico}

\author{David A.\ Ruiz-Tijerina}
\email{d.ruiz-tijerina@fisica.unam.mx}
\affiliation{Departamento de F\'isica Qu\'imica, Instituto de F\'isica, Universidad Nacional Aut\'onoma de M\'exico, Ciudad de M\'exico, C.P. 04510, M\'exico}

\begin{abstract}
We report on the theoretical electronic spectra of twisted phosphorene bilayers exhibiting moir\'e patterns, as computed by means of a continuous approximation to the moir\'e superlattice Hamiltonian. Our model is constructed by interpolating between effective $\Gamma$-point conduction- and valence-band Hamiltonians for the different stacking configurations approximately realized across the moir\'e supercell, formulated on symmetry grounds. We predict the realization of three distinct regimes for $\Gamma$-point electrons and holes at different twist angle ranges: a \emph{Hubbard} regime for small twist angles $\theta < 2^\circ$, where the electronic states form arrays of quantum-dot-like states, one per moir\'e supercell; a \emph{Tomonaga-Luttinger} regime at intermediate twist angles $2^\circ < \theta \lesssim 10^\circ$, characterized by the appearance of arrays of quasi-1D states; and, finally, a \emph{ballistic} regime at large twist angles $\theta \gtrsim 10^\circ$, where the band-edge states are delocalized, with dispersion anisotropies modulated by the twist angle. Our method correctly reproduces recent results based on large-scale \emph{ab initio} calculations at a much lower computational cost, and with fewer restrictions on the twist angles considered.
\end{abstract}

\maketitle

\section{Introduction}\label{sec:introduction}

Twisted bilayers of two-dimensional materials have quickly arisen as playgrounds for the exploration of fundamental physics, and potential platforms for novel technological applications. Unconventional superconductivity \cite{cao2018unconventional}, magnetic phenomena \cite{chen2020tunable,sharpe2019emergent} and Mott insulating phases \cite{cao2018correlated} have been observed in twisted bilayer graphene at so-called magic angles, whereas Hubbard-model physics\cite{CorrelatedTMD1,CorrelatedTMD2}, exciton miniband formation \cite{jin2019observation,alexeev2019resonantly} and confinement\cite{Seyler2019,Tran2019,Brotons2020,Soltero_2020}, and strong lattice reconstruction\cite{TMDrelax1,TMDrelax2} have been measured in semiconducting transition-metal dilcogenide (TMD) bilayers. Underlying these phenomena is the formation of moir\'e patterns: approximate superlattice structures formed by the spatial modulation of the interlayer registry across the sample plane. The long-range periodicity of the moir\'e pattern folds and couples the carrier bands, producing ultra flat minibands that promote strong correlations. Thus far, the study of moir\'e superlattices in twisted bilayers of van der Waals materials has focused strongly on graphene and TMDs. By contrast, moir\'e physics in twisted phosphorene bilayers remains largely unexplored\cite{Wallbank_2017,MoirePh2017,Fang_2019,Brooks_2020,Wang_NatCommun_2021,Wang_trilayerP}.

Here, we present a model capable of describing the electronic spectra of twisted phosphorene bilayers. Through the construction of an effective superlattice Hamiltonian that considers interlayer hybridization and intralayer energy modulation by the moir\'e potential, we compute miniband structures for electrons and holes around the $\Gamma$ point. This superlattice model is based on the so-called continuous approximation, obtained by interpolating between effective Hamiltonians for aligned phosphorene bilayers with all the different stacking configurations approximately realized inside the moir\'e supercell\cite{LandscapesAPL,LandscapesPRB}. We present a thorough, symmetry-based derivation of these effective models, which we parametrize based on density functional theory calculations for multiple aligned bilayers. We then report on the moir\'e miniband spectra of twisted phosphorene bilayers at multiple twist angles. Our results are in excellent agreement with recent large-scale \emph{ab initio} calculations\cite{MoirePh2017,Brooks_2020,Wang_trilayerP}, which are limited by computational cost to relatively small moir\'e supercells, as well as to exactly commensurate twist angles. We identify three different twist angle regimes where the main conduction and valence states take on distinct geometries. At small angles ($\theta<2^{\circ}$), we find that electrons (holes) localize at HH (HA) stacking sites [see Fig.\ \ref{fig:CoverFigure}(a)] to form mesoscale rectangular lattices, for which Hubbard model physics is anticipated. Then, at intermediate twist angles ($2<\theta\lesssim 10^{\circ}$), electrons and holes delocalize along the long axis of the unit cell to form arrays of quasi-1D states. Finally, these states become fully delocalized in two dimensions at large twist angles ($\theta\gtrsim 10^\circ$), exhibiting anisotropic dispersions efficiently modulated by the interlayer twist angle. Experimental confirmation of these qualitatively distinct limits, which we call the Hubbard, Tomonaga-Luttinger and ballistic regimes, respectively, would place phosphorene bilayers among the most versatile twistronic materials to date.

\section{Modelling approach}\label{sec:model}
\begin{figure}[t!]
    \centering
    \includegraphics[width=\columnwidth]{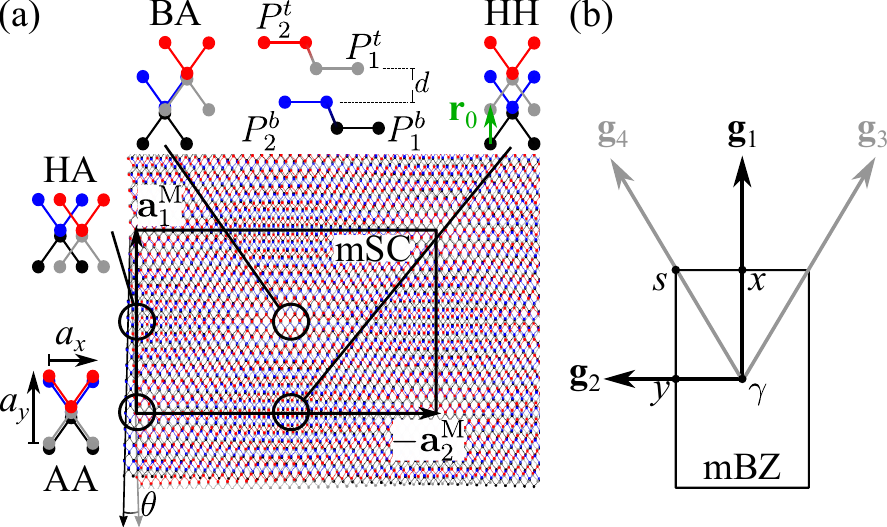}
    \caption{(a) Schematic of the moir\'e pattern appearing in a twisted phosphorene bilayer, with small twist angle $\theta=2^\circ$. The moir\'e supercell is shown, and all four high-symmetry stacking regions (see Table \ref{tab:r0}) are labeled and sketched, for clarity. (b) The moir\'e Brillouin zone corresponding to the superlattice. The first two stars of moir\'e Bragg vectors are shown, as well as the symmetry points $\gamma$, $x$, $s$, and $y$. The latter are obtained by applying the mapping \eqref{eq:moireg} the monolayer Brillouin zone symmetry points $\Gamma$, $X$, $S$, and $Y$ (not shown).}
    \label{fig:CoverFigure}
\end{figure}
The moir\'e pattern formed by a twisted phosphorene bilayer with small twist angle $\theta \ll 1$ is an approximate superlattice, as shown in Fig.\ \ref{fig:CoverFigure}(a). Locally, every region of its mSC approximates a commensurate stacking configuration between two perfectly aligned monolayers, each one uniquely determined by an in-plane offset vector $\rr_0$, representing the relative displacement of the top layer with respect to the bottom one, and a local interlayer distance $d(\rr_0)$. Figure \ref{fig:CoverFigure}(a) shows the four most highly symmetrical stacking configurations, and illustrates the offset vector $\rr_0$ in the case of HH stacking. The values taken by $\rr_0$ for all four configurations are listed in Table \ref{tab:r0}. The interlayer distance dependence on stacking will be discussed in Secs.\ \ref{sec:DFT} and \ref{sec:interpolation}.
\begin{table}[b!]
\caption{Interlayer offset vectors $\rr_0$ corresponding to the different high-symmetry stacking configurations of bilayer phosphorene.}
\begin{center}
\begin{tabular}{P{2.5cm} P{5.5cm}}
\hline\hline
Stacking & $\rr_0$\\
\hline
AA & $\boldsymbol{0}$ \\
HH & $a_y\hat{\mathbf{y}}/2$ \\
BA & $a_x\hat{\mathbf{x}}/2+a_y\hat{\mathbf{y}}/2$\\
HA & $a_x\hat{\mathbf{x}}/2$\\
 \hline\hline
\end{tabular}
\end{center}
\label{tab:r0}
\end{table}%!!!!!!!!!!!!!!!!

In reciprocal space, the $n$th Bragg vector of the moiré superlattice is $\gG_{n} = \tilde{\GG}_{t,n}' - \tilde{\GG}_{b,n}$, where $\tilde{\GG}_t$ and $\tilde{\GG}_b$ are reciprocal lattice vectors of the rotated top- and bottom layers, respectively. In terms of the reciprocal vectors $\GG_n$ of an unrotated layer, these are given by
\begin{equation}\label{eq:moireg}
    \tilde{\GG}_{b,n} = \mathcal{R}_{-\theta/2}\GG_{n}, \quad \tilde{\GG}_{t,n} = \mathcal{R}_{\theta/2}\GG_{n},
\end{equation}
where $\mathcal{R}_{\varphi}$ represents rotation by angle $\varphi$ about the $\hat{\boldsymbol{\mathrm{z}}}$ axis. In this paper, we have chosen the basis Bragg vectors
\begin{equation}\label{eq:basisBragg}
    \GG_1 = \frac{2\pi}{a_x}\hat{\mathbf{x}},\quad \GG_2=\frac{2\pi}{a_y}\hat{\mathbf{y}},
\end{equation}
with lattice constants $a_x=3.296\,{\rm \AA}$ and $a_y=4.590\,{\rm \AA}$ determined by \emph{ab initio} calculations in Sec.\ \ref{sec:DFT}, and illustrated in Fig.\ \ref{fig:CoverFigure}(a). The basis moir\'e vectors
\begin{equation}
    \gG_1 \approx \frac{2\pi}{a_x}\theta\hat{\mathbf{y}},\quad\gG_2 \approx -\frac{2\pi}{a_y}\theta\hat{\mathbf{x}},
\end{equation}
and the corresponding moir\'e Brillouin zone (mBZ) are shown in Fig.\ \ref{fig:CoverFigure}(b). The mSC vectors
\begin{equation}\label{eq:aM}
    \mathbf{a}_1^{\rm M}\approx a_x\theta^{-1}\hat{\mathbf{y}},\quad \mathbf{a}_2^{\rm M}\approx -a_y\theta^{-1} \hat{\mathbf{x}},
\end{equation}
are shown in Fig.\ \ref{fig:CoverFigure}(a).

In a large-periodicity moir\'e superlattice, where the stacking configuration varies slowly across the mSC, the low energy electronic states are well described by the so-called continuum approximation. This approach has found remarkable success in describing electrons in twisted bilayer graphene\cite{Bistritzer2011,Koshino2015A,Koshino2015B,Kim2017} and twisted homo- and heterostructures of transition-metal dichalcogenides\cite{TMDmoire1,TMDmoire2,TMDmoire3,TMDmoire4,TMDmoire5,TMDmoire6,TMDmoire7,LandscapesAPL,LandscapesPRB}. The approximation consists of treating $\rr_0$ as a vector field $\rr_0(\rr)$ defined over the continuum of points $\rr$ in the sample plane. For a two-dimensional homobilayer with twist angle $\theta\ll 1$, $\rr_0(\rr)$ is well approximated by
\begin{equation}\label{eq:r0ofr}
    \rr_0(\rr) = \theta \hat{\mathbf{z}}\times \rr.
\end{equation}
All superlattice parameters depending on $\rr_0$, such as the interlayer distance $d(\rr_0)$, can then be interpolated for all values of $\rr$ by the substitution \eqref{eq:r0ofr}, from known values at a finite number of stacking configurations.

Here, we use this approach to interpolate an effective electronic Hamiltonian for the conduction and valence bands of twisted bilayer phosphorene. To achieve this, in Sec.\ \ref{sec:DFT} we compute the band structures of multiple aligned ($\theta=0^\circ$) phosphorene bilayers at different stacking configurations using DFT. Then, in Sec.\ \ref{sec:Heff} we develop effective $\rr_0$-dependent Hamiltonians for the band-edge electrons of aligned phosphorene bilayers, using a symmetry-based approach. These models are then parametrized to match the DFT results in Sec.\ \ref{sec:parametrization}. In Sec.\ \ref{sec:interpolation}, we use Eq.\ \eqref{eq:r0ofr} to interpolate the parametrized Hamiltonians across the mSC, obtaining a continuous approximation to the superlattice Hamiltonian, which we solve numerically using zone-folding methods to compute the moir\'e mini-band structures of twisted phosphorene bilayers at multiple twist angles.

\begin{figure}[t!]
    \centering
    \includegraphics{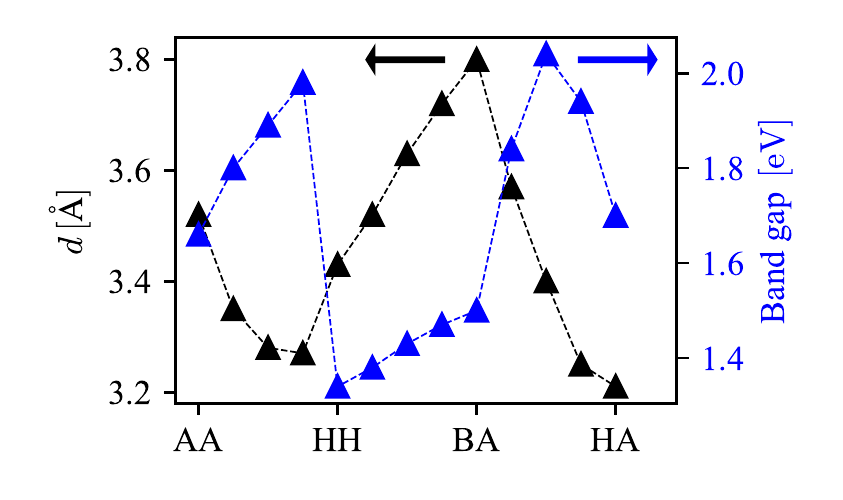}
    \caption{Interlayer distances $d$ (left) and scissor-corrected band-gap energies (right) of aligned phosphorene bilayers at the four high-symmetry stacking configurations AA, HH, BA and HA, as well as intermediate configurations. All values correspond to fully relaxed structures, as computed using PBE + Grimme-D3 DFT calculations. The obtained band gap was scissor-corrected based on the bilayer band gap for HA bilayers reported in Ref.\ \onlinecite{Castellanos_Gomez_2014}.}
    \label{fig:DFT1}
\end{figure}

\section{\emph{Ab initio} results for arbitrarily stacked phosphorene bilayers}\label{sec:DFT}
We performed DFT computations as implemented in the \textsc{Vienna Ab Initio Simulation Package}\cite{KresseHafner1993,KresseFurth1996,KRESSE199615} using the standard generalized gradient approximation, with the Perdew-Burke-Enzerhof (PBE) parametrization\cite{PBE1996}. The electronic states are treated with projector-augmented wave basis sets\cite{KresseJoubert1999}. The optimized cutoff energy for the plane wave expansion was $500\, {\rm eV}$. Since we are treating the interaction between two phosphorene layers, a dispersion-corrected van der Waals scheme is necessary. Here, we used the Grimme-D3 method, in which the dispersion coefficients are adjusted to the local geometry of the system\cite{Grimme2010}. Force and energy convergence criteria were set to 0.01 ${\rm eV/\AA}$ and $1\times 10^{-6}\,{\rm eV}$, respectively. A $\Gamma$-centered $\mathbf{k}$-points mesh of $9\times12\times1$ is used to sample the Brillouin zone (BZ) for structural optimization. For band structure calculations, a denser $\kk$-point mesh of $18\times24\times1$ was used. The same mesh was also used to achieve self consistency for the fixed interlayer distance calculations.

The monolayer phosphorene used to build the bilayer systems was fully optimized to obtain the lattice parameters reported in Sec.\ \ref{sec:model}, which are in good agreement with previous reports\cite{LiuPhosph2014,PengPhosph2014}. A vacuum gap of $\sim 20\,{\rm \AA}$ was used in the perpendicular direction to eliminate undesirable self-interactions between the phosphorene layers with their corresponding images generated by the artificial supercell periodicity. We performed computations for the four high-symmetry stacking configurations of Table \ref{tab:r0}. In addition, we considered three intermediate configurations at $\tfrac{1}{3}$, $\tfrac{1}{2}$, and $\tfrac{2}{3}$ of the path between each two subsequent high-symmetry stackings, following the path ${\rm AA}\rightarrow {\rm HH} \rightarrow {\rm BA} \rightarrow {\rm HA}$. In total, we studied 13 differently stacked bilayer systems.

Each phosphorene monolayer $\lambda$ is formed by two staggered P monolayers, $P_1^\lambda$ and $P_2^\lambda$, separated by an interlayer distance $d$ between layers $P_2^b$ and $P_1^t$, along the axis perpendicular to the sample plane [see Fig.\ \ref{fig:CoverFigure}(a)]. All models were fully optimized to obtain the interlayer distance $d$ as a function of the stacking vector $\rr_0$. Figure \ref{fig:DFT1} summarizes the main results. The shortest interlayer distance ($3.21\,{\rm \AA}$) is obtained for the HA bilayer, which constitutes the 2D building block of bulk black phosphorus\cite{Castellanos_Gomez_2014}. HA is also the most stable configuration overall, with a lower cohesive energy than AA, BA and HH stackings by $0.075 {\rm eV}$, $0.133 {\rm eV}$, and $0.069 {\rm eV}$, respectively. By contrast, BA bilayers are the least stable, and exhibit the largest interlayer distance overall ($3.80\,{\rm A}$), due to the large repulsion between P layers $P_2^b$ and $P_1^t$. These results are in good agreement with previous \emph{ab initio} results on phosphorene bilayers\cite{BilayerPhosph2014}.

It is well known\cite{DFTgap1990} that, whereas DFT calculations correctly describe the band structure topology, they do not correctly reproduce the experimental band gap. To address this issue, we implement a scissor correction, following earlier studies on semiconductors\cite{Scissor1,Scissor2, Scissor3, Scissor4, Scissor5, Scissor6, Scissor7}, using previously reported values for HA phosphorene bilayers\cite{Castellanos_Gomez_2014} that correctly reproduce experimental measurements,\footnote{The band gap reported in Ref.\ \onlinecite{Castellanos_Gomez_2014} is based on the DFT calculations that implement the Hartree-Fock corrected B3LYP functional. The reported band gap is $E_g^{\rm HA}=1.7\,{\rm eV}$. This result is consistent with the optical band gap of $1.45\,{\rm eV}$ reported in Ref.\ \onlinecite{Castellanos_Gomez_2014}, given recent calculations of $\Gamma$-point exciton binding energies in phosphorene\cite{Fabian2019,Peres2020}.}, while retaining the band gap variation as described by our DFT calculations. The scissor correction consists of a rigid translation of all conduction bands by a gap correction energy $\delta E_{\rm sc} = 1.31\,{\rm eV}$.

\begin{figure*}[t!]
    \centering
    \includegraphics{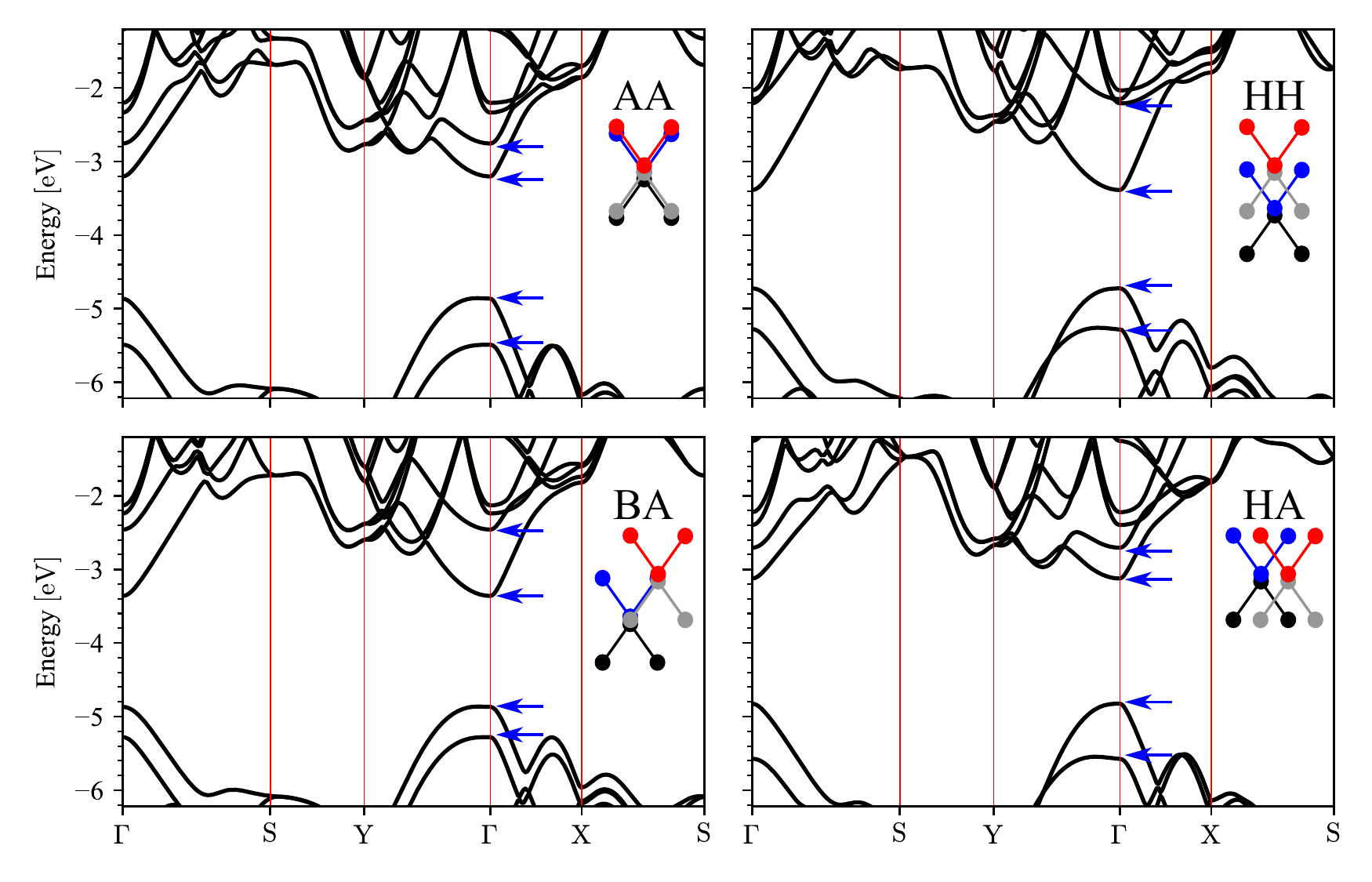}
    \caption{Band structures for AA-, HH-, BA- and HA stacked phosphorene bilayers, with the corresponding bilayer unit cells shown as insets. All four cases show a direct band gap at the $\Gamma$ point. The first two conduction- and valence subbands are indicated with blue arrows.}
    \label{fig:DFT2}
\end{figure*}

Figure \ref{fig:DFT1} shows the scissor-corrected band gap variation as a function of stacking for all 13 configurations considered. All cases exhibit semiconducting behavior, with a gap energy modulation driven by the variation of the interlayer $\pi-\pi$ interaction strength at different interlayer distances $d(\rr_0)$\cite{BilayerPhosph2014}. Note that the gap energy follows the inverse trend of the interlayer distance between configurations AA and HH, when the unit cells are translated along the $\hat{\mathbf{y}}$ direction, reaching the overall minimum gap value of $1.34\,{\rm eV}$ for HH bilayers. The band gap then grows linearly by about $200\,{\rm meV}$ along the path between configurations HH and BA, following the same trend as the interlayer distance. Finally, the band varies nonmonotonically between BA and HA, reaching an overall maximum value of $2.04\,{\rm eV}$ at the midpoint between these two configurations. Previous reports on AA, BA, HH and HA phosphorene bilayers---including the optB88-vdW functional---agree with our interlayer distances and band gap trend\cite{BilayerPhosph2014}. Our results are also consistent with data found using PBE without van der Waals interactions\cite{PBEnovdW}. Calculations using the Heyd–Scuseria–Ernzerhof functional suggest that the band gaps for these stackings are approximately $0.65 {\rm eV}$ larger than the ones obtained by our PBE + Grimme-D3 calculations, before scissor corrections. Nonetheless, the same trend is obtained for the gap variation with stacking, showing the validity of our approximation.

Figure \ref{fig:DFT2} presents detailed band structures for the four relaxed high-symmetry bilayer configurations, setting the vacuum level as a common energy reference for all cases shown. Although the four high-symmetry configurations exhibit a direct band gap at the $\Gamma$ point, analogous to the monolayer case, our DFT calculations show that four out of the nine intermediate stacking configurations deviate from this behavior, showing a valence-band maximum slightly away from $\Gamma$ along the $\overline{\Gamma \,X}$ direction, appearing approximately $115\,{\rm meV}$ above the $\Gamma$-point valence band edge and making the gap slightly indirect. Further details can be found in Appendix \ref{app:indirectgap}. It is unclear whether these features might change in a more sophisticated \emph{ab initio} scheme (\emph{e.g.}, $GW$). In the following, we shall focus on the band gap at the $\Gamma$ point when describing the different stacking regions inside a mSC formed by a twisted phosphorene bilayer. This approximation is justified if one considers that, in an actual twisted bilayer, the different stacking regions will grow or shrink according to their corresponding adhesion energies, as has been consistently observed in other 2D materials, such as graphene\cite{Grelax1,Grelax2} and transition-metal dichalcogenides\cite{TMDrelax1,TMDrelax2}. In the case at hand, BA stacking regions will shrink in favor of AA, HH and especially HA regions. The latter shall grow to form domains that occupy most of the mSC area. In turn, we expect all intermediate stacking regions to become domain walls, occupying only a small area of the mSC, and becoming less important for the description of the superlattice electronic states.

\section{Effective Hamiltonians for arbitrarily stacked phosphorene bilayers}\label{sec:Heff}

\subsection{General approach}\label{sec:GenApp}
In this section we formulate effective $\Gamma$-point Hamiltonians for the lowest two conduction- and highest two valence subbands (see Fig.\ \ref{fig:DFT2}) of an aligned phosphorene bilayer at arbitrary stacking. These models are based on same-band interlayer hybridization between monolayer states, treating interlayer conduction-valence hybridization in second-order perturbation theory, as justified by the large monolayer energy gap of $\approx 2.15\,{\rm eV}$ at the $\Gamma$ point\cite{Castellanos_Gomez_2014}. The resulting Hamiltonian for band $\alpha=c,v$ for conduction and valence, respectively, takes the form
\begin{equation}\label{eq:Hform}
    H_\alpha(\rr_0) = \begin{pmatrix}
    \varepsilon_\alpha^b(\rr_0) & T_\alpha(\rr_0)\\
    T_\alpha^*(\rr_0) & \varepsilon_\alpha^t(\rr_0)
    \end{pmatrix},
\end{equation}
where the in-plane offset vector $\rr_0$ gives the specific stacking configuration.

The hopping term $T_\alpha(\rr_0)$ corresponds to the interlayer matrix element
\begin{equation}
    T_\alpha(\rr_0)\equiv {}_b\langle \alpha,\kk | H_{\rm eff}(\rr_0) | \alpha,\kk \rangle_t,
\end{equation}
of the effective microscopic Hamiltonian $H_{\rm eff}(\rr_0)$ developed in Sec.\ \ref{sec:FullModel} below, between the Bloch states $|\alpha,\kk \rangle_t$ and $|\alpha,\kk \rangle_b$, corresponding to electrons of momentum $\kk$ and band index $\alpha$ of the top- and bottom layers, respectively. Note that we have omitted the spin index for states near the $\Gamma$ point, where spin degeneracy is guaranteed. By contrast, the top- and bottom-layer state energies $\varepsilon_\alpha^\lambda(\rr_0)$ (with $\lambda=t,\,b$ for the top and bottom layers, respectively), consists of three contributions:
\begin{equation}\label{eq:diagform}
    \varepsilon_\alpha^\lambda(\rr_0) = \varepsilon_\alpha^0 + {}_\lambda\langle \alpha,\kk | V_{\bar{\lambda}} |\alpha,\kk \rangle_\lambda + \delta \varepsilon_\alpha^\lambda (\rr_0).
\end{equation}
Here, $\varepsilon_\alpha^0$ is an $\rr_0$-independent term, containing the energy of the monolayer $\alpha$-band state at the $\Gamma$ point and $d$-dependent energy corrections, where $d$ is the interlayer distance. $V_{\bar{\lambda}}$ represents the crystal potential of the layer opposite to $\lambda$. Finally, the correction term $\delta \varepsilon_\alpha^\lambda(\rr_0)$ originates from virtual tunneling of $\alpha$-band electrons of layer $\lambda$ into the nearest bands $\beta \ne \alpha$ in the opposite layer, $\bar{\lambda}$. In the case of $\alpha=v$, we shall consider only $\beta=c$, whereas for $\alpha=c$ we shall take $\beta=v,c'$, where $c'$ is the second (monolayer) conduction band. This provides a minimal model capable of reproducing the DFT conduction- and valence subband energies, as we shall discuss in Sec.\ \ref{sec:parametrization}.

\subsection{Microscopic Hamiltonian}\label{sec:FullModel}
The microscopic Hamiltonian for a perfectly aligned, arbitrarily stacked phosphorene bilayer can be represented as\cite{LandscapesAPL,LandscapesPRB}
\begin{equation}
    H_{\rm M}(\rr,z;\rr_0) = \frac{p^2}{2m_0} + V_{t} + V_{ b},
\end{equation}
where $\mathbf{p}$ is the momentum operator, $m_0$ is the bare electron mass, and $V_{\rm \lambda}$ is the crystal potential for layer $\lambda$ centered at position $(\rr_\lambda,z_\lambda)$, with $\rr_\lambda$ an in-plane vector. In terms of $\rr_0$, $\rr_t=\rr_0/2$ and $\rr_b=-\rr_0/2$. Placing the coordinate origin on the middle plane between the two layers, we can write $z_{t/b}=\pm d(\rr_0)/2$, with $d$ the interlayer distance.

Let $| \alpha,\kk \rangle_{\lambda}$ be the Bloch wave function for the electronic state with wave vector $\kk$ and band index $\alpha$ of the isolated monolayer $\lambda$, defined such that
\begin{equation}
    \left[\frac{p^2}{2m_0} + V_{\lambda} \right]|\alpha,\kk \rangle_\lambda = \varepsilon_{\alpha}(\kk)|\alpha,\kk \rangle_\lambda,
\end{equation}
with $\varepsilon_{\alpha}(\kk)$ the corresponding band dispersion. The matrix elements of $H_{\rm M}$ are
\begin{equation}\label{eq:HMelems}
\begin{split}
    &{}_{\lambda}\langle  \alpha',\kk' | H_{\rm M} | \alpha,\kk  \rangle_{\lambda} = \delta_{\alpha',\alpha}\delta_{\kk',\kk}\varepsilon_{\alpha}(\kk)\\
    &\qquad\qquad\qquad\qquad\quad+ {}_{\lambda}\langle \alpha',\kk' | V_{\bar{\lambda}} | \alpha,\kk \rangle_{\lambda},\\
    &{}_{t}\langle  \alpha',\kk' | H_{\rm M} | \alpha,\kk  \rangle_{b} =  \left[\varepsilon_{\alpha'}(\kk')+\varepsilon_{\alpha}(\kk) \right]S_{\kk',\kk}^{\alpha',\alpha}-K_{\kk',\kk}^{\alpha',\alpha},
\end{split}
\end{equation}
where $S_{\kk',\kk}^{\alpha',\alpha}$ is the interlayer overlap matrix element
\begin{equation}
    S_{\kk',\kk}^{\alpha',\alpha}\equiv  {}_{t}\langle \alpha',\kk' | \alpha,\kk \rangle_{b},
\end{equation}
and $K_{\kk',\kk}^{\alpha',\alpha}$ is the kinetic energy interlayer matrix element
\begin{equation}
    K_{\kk',\kk}^{\alpha',\alpha} \equiv  \frac{{}_{t}\langle \alpha',\kk'    |p^2| \alpha,\kk \rangle_{b}}{2m_0}.
\end{equation}
Writing a general eigenstate of the system in the form
\begin{equation}
    |\Psi \rangle = \sum_{\alpha,\kk}\left(A_{\alpha,\kk}^{\Psi}|\alpha,\kk \rangle_{b} + B_{\alpha,\kk}^{\Psi}|\alpha,\kk \rangle_{t}\right),
\end{equation}
we obtain the generalized eigenvalue problem
\begin{equation}\label{eq:GenEig}
    H_0\Psi = (\mathbbm{1}+S)E\Psi,
\end{equation}
with the column vector of coefficients $\Psi$ defined as
\begin{equation}
    \Psi = \begin{pmatrix} \{ A_{\alpha,\kk}^\Psi \} \\ \{ B_{\alpha,\kk}^\Psi \}  \end{pmatrix},
\end{equation}
and $H_0$ containing all matrix elements of $H_{\rm M}$, except for those involving the overlap matrix $S$. Following Ferreira \emph{et al.} \cite{LandscapesAPL} and Magorrian \emph{et al.} \cite{LandscapesPRB}, we transform Eq.\ \eqref{eq:GenEig} into a proper Schr\"odinger equation by applying the unitary transformation $\mathcal{U}=(\mathbbm{1}+S)^{-1/2}$, resulting in the effective Hamiltonian
\begin{equation}\label{eq:Heff}
    H_{\rm eff} \equiv \mathcal{U}^{-1}H_{0}\mathcal{U}= H_0 - \frac{\{S,\,H_0 \}}{2} +  \mathcal{O}\{S^2\},
\end{equation}
where $\{S,\,H_0 \}$ is the anti-commutator of matrices $S$ and $H_0$. Since the states $|\alpha,\kk \rangle_\lambda$ decay exponentially in the out-of-plane direction, away from $z_\lambda$, the matrix elements of $S$ are exponentially suppressed by the interlayer distance. This justifies treating $S$ as a perturbation, and truncating the expansion \eqref{eq:Heff} at first order in $S$. Note, however, that $H_0$ contains interlayer matrix elements [see\ Eq.\ \eqref{eq:HMelems}], which are also of order $S$. Up to first order in $S$, the intra- and interlayer matrix elements of $H_{\rm eff}$ are
\begin{equation}\label{eq:Heffelems}
\begin{split}
    {}_{\lambda}\langle \alpha',\kk' | H_{\rm eff} |\alpha,\kk \rangle_{\lambda} =&  {}_{\lambda}\langle \alpha',\kk' | V_{\bar{\lambda}} |\alpha,\kk \rangle_{\lambda},\\
    {}_{t}\langle \alpha',\kk' | H_{\rm eff} |\alpha,\kk \rangle_{b} =& \frac{\varepsilon_{\alpha'}(\kk')+\varepsilon_{\alpha}(\kk)}{2}S_{\kk',\kk}^{\alpha',\alpha}\\
    & - K_{\kk',\kk}^{\alpha',\alpha}.
\end{split}
\end{equation}

One of the useful features of the above formulation for the homobilayer Hamiltonian is that all symmetry constraints are encoded in the wave functions and crystal potentials. The former may be Fourier expanded as
\begin{equation}\label{eq:Fourier}
    \psi_{\alpha,\kk}^\lambda(\rr,z) \equiv \langle \rr,z | \alpha,\kk \rangle_\lambda = \sum_{\GG}\frac{e^{i(\GG+\kk)\cdot(\rr-\rr_\lambda)}}{\sqrt{N}}u_\alpha^\lambda(\GG+\kk,z),
\end{equation}
where $\GG$ are the bilayer Bragg vectors, $N$ the number of unit cells in the sample, and the Fourier coefficients $u_\alpha^\lambda$ are decaying functions of $|z-z_\lambda|$. Then, the interlayer overlap- and kinetic energy matrix elements can be written as
\begin{subequations}
\begin{equation}\label{eq:overlap}
\begin{split}
    S_{\kk',\kk}^{\alpha',\alpha}=&\sum_{\GG,\GG'}\delta_{\GG'+\kk',\GG+\kk}e^{i(\GG'\cdot\rr_t - \GG\cdot\rr_b)}\\
    &\times\int dz\, u_{\alpha'}^{t\,*}(\GG'+\kk',z)u_\alpha^b(\GG+\kk,z),
\end{split}
\end{equation}
\begin{equation}\label{eq:pxpect}
\begin{split}
    &K_{\kk',\kk}^{\alpha',\alpha}= -\sum_{\GG,\GG'}\frac{\hbar^2}{2m_0}\delta_{\GG'+\kk',\GG+\kk}e^{i(\GG'\cdot\rr_t - \GG\cdot\rr_b)}\\
    &\,\times\left( \left|\GG+\kk \right|^2\int d^2r\,u_{\alpha'}^{t*}(\GG'+\kk',z)u_\alpha^b(\GG+\kk,z) \right.\\
    &\,+ \left. \int dz\, \left[\partial_z u_{\alpha'}^{t*}(\GG'+\kk',z)\right] \left[\partial_z u_\alpha^b(\GG+\kk,z)\right]\right).
\end{split}
\end{equation}
\end{subequations}
\begin{table}[t!]
	\caption{
		{Character table for point group $D_{2h}$, describing the symmetry properties of the monolayer phosphorene bands \cite{LiApplebaum2014}. \label{tab:character}
		}
	}
	\begin{tabular}{P{0.82cm} | P{0.82cm} P{0.82cm} P{0.82cm} P{0.82cm} P{0.82cm} P{0.82cm} P{0.82cm} P{0.82cm}}
		\hline\hline
		\, & $E$ & $\tilde{C}_{2,\uvec{z}}$ & $\tilde{C}_{2,\uvec{y}}$ & $C_{2,\uvec{x}}$ & $\mathcal{I}$ & $\sigma_{xy}$ & $\sigma_{zx}$ & $\sigma_{yz}$\\
		\hline
		$A_g$   & $+1$ & $+1$ & $+1$ & $+1$ & $+1$ & $+1$ & $+1$ & $+1$ \\
		$B_{1g}$& $+1$ & $+1$ & $-1$ & $-1$ & $+1$ & $+1$ & $-1$ & $-1$ \\
		$B_{2g}$& $+1$ & $-1$ & $+1$ & $-1$ & $+1$ & $-1$ & $+1$ & $-1$ \\
		$B_{3g}$& $+1$ & $-1$ & $-1$ & $+1$ & $+1$ & $-1$ & $-1$ & $+1$ \\
		$A_u$   & $+1$ & $+1$ & $+1$ & $+1$ & $-1$ & $-1$ & $-1$ & $-1$ \\
		$B_{1u}$& $+1$ & $+1$ & $-1$ & $-1$ & $-1$ & $-1$ & $+1$ & $+1$ \\
		$B_{2u}$& $+1$ & $-1$ & $+1$ & $-1$ & $-1$ & $+1$ & $-1$ & $+1$ \\
		$B_{3u}$& $+1$ & $-1$ & $-1$ & $+1$ & $-1$ & $+1$ & $+1$ & $-1$ \\
		\hline\hline
	\end{tabular}
\end{table}
At this point, we introduce two approximations: first, we approximate the Fourier series \eqref{eq:Fourier} by its dominant terms, corresponding to the first three stars of Bragg vectors:
\begin{equation}\label{eq:3stars}
\begin{split}
    &\GG_0=\mathbf{0},\\
    &\GG_{\pm1} = \pm\tfrac{2\pi}{a_x},\,\GG_{\pm2}=\pm \tfrac{2\pi}{a_y},\\
    &\GG_{\pm3} = \pm \left(\GG_1-\GG_2 \right),\,\GG_{\pm4} = \pm \left(\GG_1+\GG_2  \right),\\
    &\GG_{\pm5} = \pm 2 \GG_1,\,\GG_{\pm6}=\pm 2\GG_2.
\end{split}
\end{equation}
Second, we assume that the functions $u_{\alpha}(\GG+\kk,z)$ vary slowly with wave vector, such that for $\kk$ close enough to the $\Gamma$ point we may approximate $u_{\alpha}(\GG+\kk,z)\approx u_{\alpha}(\GG,z)$. Finally, since we are concerned exclusively with wave vectors $\kk$ and $\kk'$ much smaller than any reciprocal lattice vector, we may write
\begin{equation}
    \delta_{\kk-\kk',\GG_{n'}-\GG_n} = \delta_{\kk,\kk'}\delta_{n,n'}.
\end{equation}
Then, Eqs.\ \eqref{eq:overlap} and \eqref{eq:pxpect} simplify to
\begin{subequations}
\begin{equation}\label{eq:overlap_simplified}
    S_{\kk',\kk}^{\alpha',\alpha} \approx \delta_{\kk,\kk'}\sum_{n=-6}^6e^{i\GG_n\cdot\rr_0}S_{n,n}^{\alpha',\alpha},
\end{equation}
\begin{equation}\label{eq:pxpect_simplified}
K_{\kk',\kk}^{\alpha',\alpha}  \approx -\delta_{\kk,\kk'}\sum_{n=-6}^6\frac{\hbar^2e^{i\GG_n\cdot\rr_0}}{2m_0}\left( G_n ^2S_{n,n}^{\alpha',\alpha} + D_{n,n}^{\alpha',\alpha}\right),
\end{equation}
\end{subequations}
where we have defined
\begin{equation}
\begin{split}
    S_{n',n}^{\alpha',\alpha} \equiv& \int dz\, u_{\alpha'}^{t*}(\GG_{n'},z)u_\alpha^b(\GG_n,z),\\
    D_{n',n}^{\alpha',\alpha} \equiv& \int dz\, \left[\partial_z u_{\alpha'}^{t*}(\GG_{n'},z)\right]\left[ \partial_z u_\alpha^b(\GG_n,z)\right].
\end{split}
\end{equation}

\subsection{Symmetry properties of the Bloch functions}
As discussed by Li and Appelbaum\cite{LiApplebaum2014}, the monolayer phosphorene crystal structure is described by the point symmetry  group $D_{2h}$, with modified symmetry operations $\tilde{C}_{2,\hat{\mathbf{z}}} = \mathcal{T}C_{2,\hat{\mathbf{z}}}$ and $\tilde{C}_{2,\hat{\mathbf{y}}} = \mathcal{T}C_{2,\hat{\mathbf{y}}}$, where $C_{m,\hat{\mathbf{n}}}$ represents a rotation by $2\pi/m$ about the axis $\hat{\mathbf{n}}$, and $\mathcal{T}$ is an in-plane translation by the unit cell vector $\mathbf{r}_{\mathcal{T}}=\tfrac{a_x}{2}\uvec{x} + \tfrac{a_y}{2}\uvec{y}$. Accordingly, all $\Gamma$-point states transform under symmetry operations like one of the group's irreducible representations (irreps). The transformation rules are especially simple for group $D_{2h}$, which contains only one-dimensional irreps, and take the form
\begin{equation}\label{eq:forinstance}
    \mathcal{D}\psi_{\alpha,\mathbf{0}}^\lambda(\rr,z) = \phi(I_\alpha,\mathcal{D}) \psi_{\alpha,\mathbf{0}}^\lambda(\rr,z),
\end{equation}
where $\mathcal{D}\in D_{2h}$; $I_\alpha$ is the irreducible representation of group $D_{2h}$ corresponding to band $\alpha$; and $\phi(I_\alpha,\mathcal{D})$ is the character of $I_\alpha$ under the symmetry operation $\mathcal{D}$ (see Table \ref{tab:character}). Note that Eq.\ \eqref{eq:forinstance} applies only to states exactly at the $\Gamma$ point.

We may obtain the symmetry properties of the Fourier coefficients $u_\alpha^\lambda(\GG,z)$ by inverse-Fourier transforming Eq.\ \eqref{eq:Fourier}:
\begin{equation}\label{eq:inverseFourier}
\begin{split}
    u_\alpha^\lambda(\GG,z)=\int d^2r\,\frac{e^{-i\GG\cdot \rr}}{\sqrt{N}}\psi_{\alpha,\mathbf{0}}^\lambda(\rr,z).
\end{split}
\end{equation}
Combining Eqs.\ \eqref{eq:forinstance} and \eqref{eq:inverseFourier} for $\mathcal{D}=\tilde{C}_{2,\hat{\mathbf{z}}}$ we obtain
\begin{widetext}
\begin{equation}\label{eq:CaseC_2z}
\begin{split}
    u_\alpha^\lambda(\GG,z)=&\int d^2r\,\frac{e^{-i\GG\cdot \rr}}{\sqrt{N}}\left[ \frac{\tilde{C}_{2,\uvec{z}}}{\phi(I_\alpha,\tilde{C}_{2,\uvec{z}})}\psi_{\alpha,\mathbf{0}}^\lambda(\rr,z) \right] = \frac{1}{\phi(I_\alpha,\tilde{C}_{2,\uvec{z}})}\int d^2r\,\frac{e^{-i\GG\cdot \rr}}{\sqrt{N}}\psi_{\alpha,\mathbf{0}}^\lambda(\tilde{C}_{2,\uvec{z}}^{-1}\rr,z)\\
    =& \frac{1}{\phi(I_\alpha,\tilde{C}_{2,\uvec{z}})}\int d^2r\,\frac{e^{-i\GG\cdot \rr}}{\sqrt{N}}\psi_{\alpha,\mathbf{0}}^\lambda(C_{2,\uvec{z}}^{-1}\mathcal{T}^{-1}\rr,z)=\frac{e^{-i\GG\cdot \rr_{\mathcal{T}}}}{\phi(I_\alpha,\tilde{C}_{2,\uvec{z}})}\int d^2\bar{r}\,\frac{e^{-i\GG\cdot \bar{\rr}}}{\sqrt{N}}\psi_{\alpha,\mathbf{0}}^\lambda(C_{2,\uvec{z}}^{-1}\bar{\rr},z)\\
    =& \frac{e^{-i\GG\cdot \rr_{\mathcal{T}}}}{\phi(I_\alpha,\tilde{C}_{2,\uvec{z}})}\int d^2\bar{r}\,\frac{e^{-iC_{2,\uvec{z}}^{-1}\GG\cdot\RR} }{\sqrt{N}}\psi_{\alpha,\mathbf{0}}^\lambda(\RR,z) = \frac{e^{-i\GG\cdot \rr_{\mathcal{T}}}}{\phi(I_\alpha,\tilde{C}_{2,\uvec{z}})} u_\alpha^\lambda (C_{2,\uvec{z}}^{-1}\GG,z),
\end{split}
\end{equation}
\end{widetext}
where we defined $\bar{\rr}=\mathcal{T}^{-1}\rr=\rr-\rr_{\mathcal{T}}$ and $\RR=C_{2,\uvec{z}}^{-1}\bar{\rr}$. Since in two dimensions rotations by $\pi$ are equivalent to in-plane inversion, Eq.\ \eqref{eq:CaseC_2z} can be rewritten as
\begin{equation}\label{eq:ConstraintC_2z}
    u_{\alpha}^\lambda(-\GG,z)=\phi(I_\alpha,\tilde{C}_{2,\uvec{z}})e^{i\GG\cdot \rr_{\mathcal{T}}} u_\alpha^\lambda (\GG,z).
\end{equation}
A similar analysis of the symmetry operations $\tilde{C}_{2,\uvec{y}}$, $C_{2,\uvec{x}}$ and inversion $\mathcal{I}$ yields the constraints
\begin{subequations}\label{eqs:Constraints}
\begin{equation}\label{eq:ConstraintC_2y}
    u_\alpha^\lambda(\sigma_{yz}\GG,-z)=\phi(I_\alpha,\tilde{C}_{2,\uvec{y}})e^{i\GG\cdot\rr_{\mathcal{T}}} u_\alpha^\lambda(\GG,z),
\end{equation}
\begin{equation}\label{eq:ConstraintC_2x}
    u_\alpha^\lambda(\sigma_{zx}\GG,-z)=\phi(I_\alpha,C_{2,\uvec{x}}) u_\alpha^\lambda(\GG,z),
\end{equation}
\begin{equation}\label{eq:ConstraintI}
    u_\alpha^\lambda(-\GG,-z)=\phi(I_\alpha,\mathcal{I}) u_\alpha^\lambda(\GG,z),
\end{equation}
\end{subequations}
respectively, with the mirror reflections $\sigma_{yz}\GG=-G_x\uvec{x} + G_y\uvec{y}$ and $\sigma_{zx}\GG=G_x\uvec{x} - G_y\uvec{y}$. Note that Eqs.\ \eqref{eq:ConstraintC_2z} and \eqref{eq:ConstraintC_2y}-\eqref{eq:ConstraintI} apply also to $\partial_z u_\alpha(\GG,z)$, appearing in $K_{\kk',\kk}^{\alpha',\alpha}$.

In Appendix \ref{app:symmetry}, we show how Eqs.\ \eqref{eq:ConstraintC_2z}--\eqref{eq:ConstraintI} constrain the different Fourier components $u_\alpha^\lambda(\GG_n)$ of the monolayer phosphorene conduction ($\alpha=c,\,I_c=B_{1u}$) and valence ($\alpha=v,\,I_v=B_{3g}$) bands, for the Bragg vectors of Eq.\ \eqref{eq:3stars}. The results are summarized in Eq.\ \eqref{eqs:all_constraints}. Combined with \eqref{eq:overlap_simplified}, \eqref{eq:pxpect_simplified} and \eqref{eq:Heffelems}, these constraints give the interlayer tunneling matrix elements 
\begin{subequations}\label{eqs:Melems}
\begin{equation}\label{eqs:MelemsVB}
\begin{split}
    T_v^*(\rr_0) \equiv {}_{t}\langle v,\kk | H_{\rm eff} |v,\kk \rangle_b =\sum_{n=1}^6t_v^{(n)}\cos{(\GG_n\cdot\rr_0)},
\end{split}
\end{equation}
\begin{equation}\label{eqs:MelemsCB}
    T_c^*(\rr_0) \equiv {}_{t}\langle c,\kk | H_{\rm eff} |c,\kk \rangle_b =\sum_{n=0}^6t_c^{(n)}\cos{(\GG_n\cdot\rr_0)},
\end{equation}
\begin{equation}
    t_v^{(4)}=t_v^{(3)},\,t_c^{(4)}=t_c^{(3)},\,t_{v}^{(5)}=t_v^{(0)}=t_c^{(1)}=0,
\end{equation}
\end{subequations}
with the definitions
\begin{equation}\label{eq:tvn}
    t_{\alpha}^{(n)} \equiv 2\left[ \frac{\varepsilon_{\alpha}(\Gamma) + \varepsilon_{\alpha}(\Gamma)}{2} + \frac{\hbar^2 G_n^2}{2m_0}  \right]S_{n,n}^{\alpha,\alpha} + \frac{\hbar^2D_{n,n}^{\alpha,\alpha}}{2m_0}.
\end{equation}

A similar analysis (see Appendix \ref{app:crystalpotential}) yields the intralayer matrix elements
\begin{subequations}\label{eq:Vall}
\begin{equation}\label{eq:Vbot}
\begin{split}
    {}_{b}\langle \alpha,\kk |V_{t}|\alpha,\kk \rangle_{b} =& v_\alpha^{(0)} - v_\alpha^{(2)}\sin{\left( \GG_2\cdot\rr_0\right)}\\ &+ \sum_{n=3}^6 v_\alpha^{(n)} \cos{\left(\GG_n\cdot\rr_0 \right)},
\end{split}
\end{equation}
\begin{equation}\label{eq:Vtop}
\begin{split}
    {}_{t}\langle \alpha,\kk |V_b|\alpha,\kk \rangle_{t} =& v_\alpha^{(0)} + v_\alpha^{(2)}\sin{\left( \GG_2\cdot\rr_0\right)}\\ &+ \sum_{n=3}^6 v_\alpha^{(n)} \cos{\left(\GG_n\cdot\rr_0 \right)},
\end{split}
\end{equation}
\begin{equation}
    v_\alpha^{(4)}=v_\alpha^{(3)},
\end{equation}
\end{subequations}
obtained from Eq.\ \eqref{eq:AppV} by substituting $\rr_{t/b} = \pm\rr_0/2$.

Finally, the virtual tunneling corrections $\delta \varepsilon_\alpha^\lambda (\rr_0)$ are obtained by considering first-order interlayer tunneling between band $\alpha$ and nearby bands $\beta$, as described in Sec.\ \ref{sec:GenApp}, and then projecting out band $\beta$ up to second order in perturbation theory by means of L\"owdin's partitioning method\cite{WinklerBook}. The details are discussed in Appendix \ref{app:virtual}. Here, we merely state the result ($\alpha=c,v$):
\begin{equation}\label{eq:dE}
\begin{split}
    \delta \varepsilon_\alpha^b (\rr_0) = \delta \varepsilon_\alpha^t& (\rr_0) = w_\alpha^{(0)}+\sum_{n=1}^6w_\alpha^{(n)}\cos{\left(\GG_n\cdot\rr_0 \right)},\\
    &w_\alpha^{(5)}=0,\, w_\alpha^{(4)}=w_\alpha^{(3)}.
\end{split}
\end{equation}
Putting together Eqs.\ \eqref{eq:Vall} and \eqref{eq:dE} and comparing with Eq.\ \eqref{eq:diagform} gives
\begin{subequations}\label{eq:Ediag}
\begin{equation}
\begin{split}
    \varepsilon_\alpha^{t/b}(\rr_0) =& \varepsilon_\alpha^{(0)}  \mp \phi_\alpha^{(2)}\sin{\left(\GG_2\cdot\rr_0 \right)}\\
    & + \sum_{n=1}^6 \varepsilon_\alpha^{(n)}\cos{\left(\GG_n\cdot\rr_0 \right)},
\end{split}
\end{equation}
\begin{equation}
    \varepsilon_\alpha^{(n)} = v_\alpha^{(n)} + w_\alpha^{(n)},
\end{equation}
\begin{equation}
    \varepsilon_\alpha^{(4)} = \varepsilon_\alpha^{(3)},\quad \varepsilon_\alpha^{(5)} = 0.
\end{equation}
\end{subequations}

This completes the derivation of all terms defining the Hamiltonian \eqref{eq:Hform}. Although Eqs.\ \eqref{eqs:MelemsVB}, \eqref{eqs:MelemsCB}, \eqref{eq:Vbot}, \eqref{eq:Vtop} and \eqref{eq:dE} can ultimately be expressed in terms of microscopic quantities, our strategy will consist of fitting all parameters $\{t_{\alpha}^{(n)}\},\,\{v_\alpha^{(n)}\},\,\{w_\alpha^{(n)}\}$ to first principles results, as described in Sec.\ \ref{sec:parametrization} below.

\subsection{\emph{Ab initio} parametrization of the effective model}\label{sec:parametrization}

\begin{figure}[t!]
    \centering
    \includegraphics{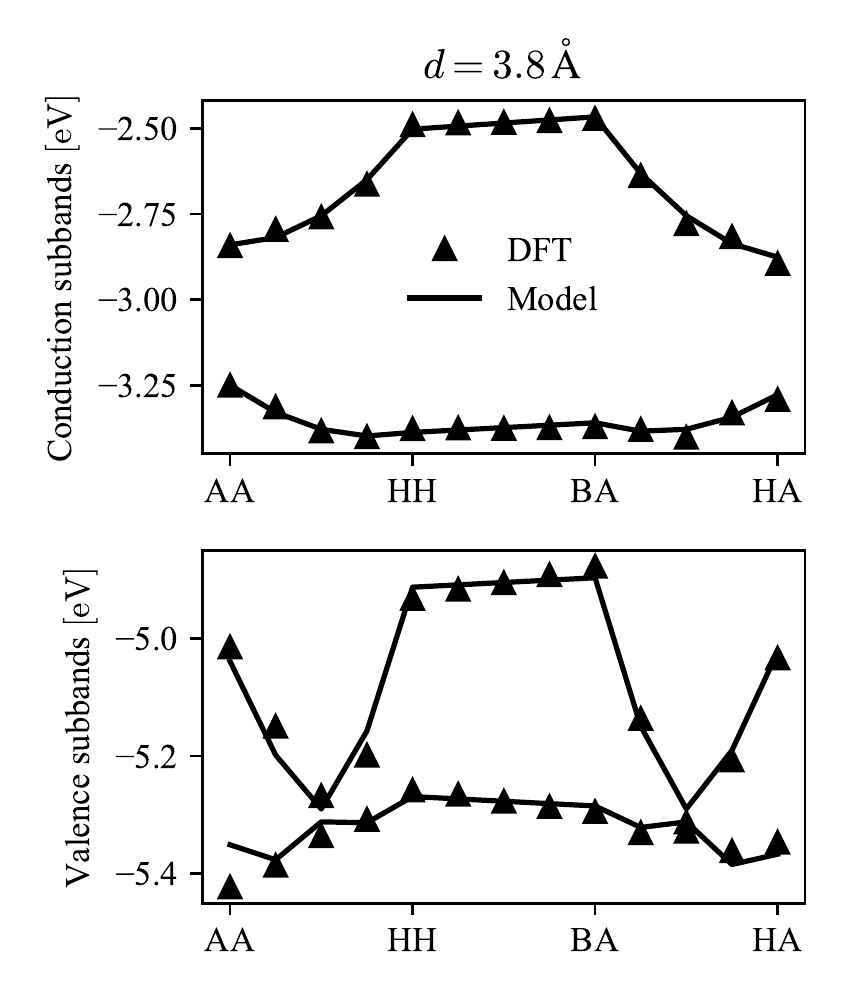}
    \caption{Stacking dependence of the bottom two conduction- and top two valence subband energies at the $\Gamma$ point, for aligned phosphorene bilayers with a fixed interlayer distance of $3.8\,\mathrm{\AA}$. Triangles represent the (scissor-corrected) DFT-computed energies, whereas solid lines show the energies obtained from the effective models \eqref{eq:Hform}, with parameters fitted to the DFT values.}
    \label{fig:Subbands_3_8A}
\end{figure}

The tunneling coefficients $t_{\alpha}^{(n)}$ and potential energies $\varepsilon_\alpha^{(n)}$ and $\phi_\alpha^{(2)}$ appearing in Eqs.\ \eqref{eqs:Melems} and \eqref{eq:Ediag}, respectively, were fitted to reproduce the conduction- and valence subband (see Fig.\ \ref{fig:DFT2}) energies obtained from scissor-corrected DFT calculations for multiple stacking configurations $\rr_0$, at fixed interlayer distance $d$. Figure \ref{fig:Subbands_3_8A} shows a comparison between the DFT-computed energies, and the energies obtained from the effective models \eqref{eq:Hform} with parameters fitted to the DFT data, for fixed interlayer distance $d=3.8\,\mathrm{\AA}$. The fitting procedure was repeated for $d=3.7$ and $3.6\,\mathrm{\AA}$. In all cases, we found that good agreement with the DFT energies can be obtained setting $\phi_\alpha^{(2)}=\varepsilon_\alpha^{(1)}=t_c^{(5)}=\varepsilon_v^{(3)}=0$. Note that $\phi_\alpha^{(2)}=0$ suggests the absence of an out-of-plane ferroelectric effect, such as that observed in transition-metal dichalcogenide homo- and heterobilayers\cite{Weston_2022}, in agreement with recent \emph{ab initio} results\cite{Nanshu_2020}.

The $d$ dependence of each remaining model parameter was then fitted as $A(d) = A_0e^{-q(d-d_0)}$, taking $d_0=3.49\,{\rm \AA}$ as a reference interlayer distance (see Sec.\ \ref{sec:interpolation} below). In the limit of $d \gg d_0$, corresponding to two decoupled bilayers, all model parameters should vanish, with the exception of $\varepsilon_\alpha^{(0)}$, which should converge to the monolayer $\alpha$-band edge energies, $\varepsilon_\alpha^{\rm ML}$. Therefore, in the following we rewrite:
\begin{equation}
    \varepsilon_\alpha^{(0)} = \varepsilon_{\alpha}^{\rm ML} + \delta \varepsilon_{\alpha}^{{0}}(d),
\end{equation}
with the scissor-corrected DFT values $\varepsilon_v^{\rm ML} = -5.012\,{\rm eV}$ and $\varepsilon_c^{\rm ML} = \varepsilon_v^{\rm ML} + \delta E_{\rm sc} = -3.702\,{\rm eV}$.
All $d$-dependent model parameters are summarized in Table \ref{tab:ModelParameters}.

\begin{table}[h!]
\caption{Interlayer distance dependence of all non-zero effective-model parameters. Each parameter $A$ was fitted to DFT data as $A=A_0e^{-q(d-d_0)}$.}
\begin{center}
\begin{tabular}{P{0.8cm}  Q{0.5cm}@{.}R{0.9cm}  Q{0.6cm}@{.}R{0.8cm}  P{0.8cm}  Q{0.5cm}@{.}R{0.9cm}  Q{0.6cm}@{.}R{0.8cm} }
\hline\hline
$A$ & \multicolumn{2}{c}{$A_0\,[\mathrm{eV}]$} & \multicolumn{2}{c}{$q\,[\mathrm{\AA}^{-1}]$} & $A$ & \multicolumn{2}{c}{$A_0\,[\mathrm{eV}]$} & \multicolumn{2}{c}{$q\,[\mathrm{\AA}^{-1}]$}\\
\hline\hline
$t_c^{(0)}$ & $0$&$384$ & $0$&$61$ & $\delta\varepsilon_c^{(0)}$ & $0$&$753$ & $0$&$37$  \\
$t_c^{(2)}$ & $-0$&$185$ & $1$&$37$ & $\varepsilon_c^{(2)}$ & $-0$&$094$ & $1$&$17$  \\
$t_c^{(3)}$ &  $0$&$003$ & $2$&$72$ & $\varepsilon_c^{(3)}$ & $-0$&$011$ & $1$&$11$  \\
$t_c^{(6)}$ &  $0$&$013$ & $2$&$50$ & $\varepsilon_c^{(6)}$ &  $0$&$058$ & $1$&$49$  \\
$t_v^{(1)}$ &  $0$&$023$ & $1$&$11$ & $\delta\varepsilon_v^{(0)}$ & $-0$&$209$ & $0$&$00$  \\
$t_v^{(2)}$ &  $0$&$266$ & $1$&$34$ & $\varepsilon_v^{(2)}$ & $-0$&$068$ & $0$&$89$  \\
$t_v^{(3)}$ & $-0$&$010$ & $2$&$90$ & $\varepsilon_v^{(6)}$ &  $0$&$136$ & $2$&$08$  \\
$t_v^{(6)}$ & $-0$&$022$ & $2$&$25$ & \, \\
\hline\hline
\end{tabular}
\end{center}
\label{tab:ModelParameters}
\end{table}%!!!!!!!!!!!!!!!!

\section{Continuous model for the moir\'e superlattice}\label{sec:interpolation}
\begin{figure}[h!]
    \centering
    \includegraphics[width=0.9\columnwidth]{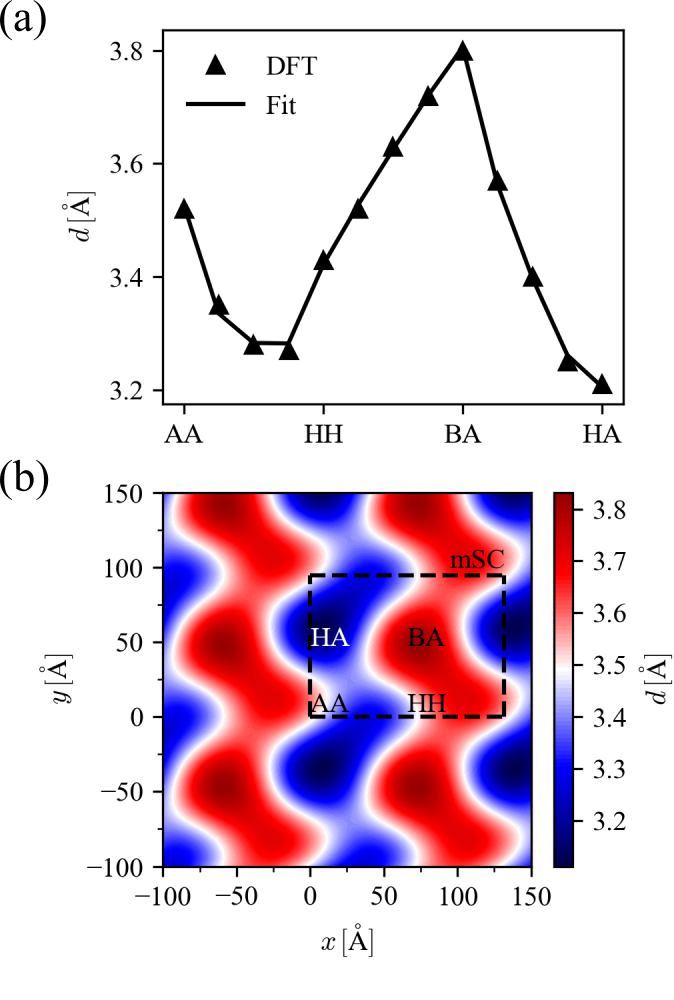}
    \caption{(a) Stacking-dependent interlayer distance, $d$. DFT results for relaxed structures are shown with triangles, whereas the solid line represents the fitting function Eq.\ \eqref{eq:d_interpolation} with the parameters shown in Table \ref{tab:d_fit}. (b) Interlayer distance as a function of position in a twisted phosphorene bilayer with $\theta=2^\circ$, obtained from Eq.\ \eqref{eq:d_interpolation} by replacing $\rr_0=\theta\hat{\mathbf{z}}\times \rr$. The color white corresponds to the reference interlayer distance $d_0=3.49\,{\rm \AA}$. The moir\'e supercell and different stacking regions are labeled in the figure.}
    \label{fig:d_interpolation}
\end{figure}

As described in Sec.\ \ref{sec:model}, the continuum approximation to the moir\'e superlattice Hamiltonian is obtained by substituting $\rr_0(\rr)= \theta \hat{\mathbf{z}}\times\rr$ into Eq.\ \eqref{eq:Hform}, with $\rr$ the in-plane position in the moir\'e superlattice, measured with respect to the center of a given AA-stacked region. Note, however, that the resulting model contains the interlayer distance $d$ as a fixed parameter. To take into account out-of-plane lattice relaxation, \emph{i.e.}, the fact that the interlayer distance will vary across the mSC according to the local stacking, we interpolate the DFT values for the interlayer distance reported in Fig.\ \ref{fig:DFT1} by the Fourier series
\begin{equation}\label{eq:d_interpolation}
    d(\rr_0) = d_0 + \sum_{n=1}^{N}\left[ d_n^s \cos{\left(\GG_n\cdot\rr_{0} \right)} + d_n^a \sin{\left(\GG_n\cdot\rr_{0} \right)}\right],
\end{equation}
where the constant $d_0$ was chosen as a reference interlayer distance when fitting the model parameters reported in Table \ref{tab:ModelParameters}. Good agreement between \eqref{eq:d_interpolation} and the DFT results is obtained for $N=4$, with the fitting parameters reported in Table \ref{tab:d_fit}, as shown in Fig.\ \ref{fig:d_interpolation}. Then, the full spatial dependence of each parameter $A$ in Table \ref{tab:ModelParameters} takes the form
\begin{equation}
\begin{split}
    &A(\rr)=A_0e^{-q[d(\rr)-d_0]}\\
    =&A_0e^{\left(-q\sum_{n=1}^4\left[d_n^s\cos{(\gG_n\cdot\rr)} + d_n^a\sin{(\gG_n\cdot\rr)} \right]\right)}.
\end{split}
\end{equation}
where we have noted that, upon the substitution $\rr_0=\theta\hat{\mathbf{z}}\times \rr$,
\begin{equation}
    \GG_n\cdot\rr_0 \approx -\gG_n\cdot\rr,    
\end{equation}
for $\theta \ll 1$. In practice, we utilize a first-order expansion of the exponential
\begin{equation}
\begin{split}
    A(\rr)=&A_0e^{-q[d(\rr)-d_0]}\approx A_0\left(1-q[d(\rr)-d_0] \right)\\
    \approx& A_0-qA_0\sum_{n=1}^4d_n^s\cos{\left(\gG_n\cdot\rr \right)}\\
    &\quad\,+qA_0\sum_{n=1}^4d_n^a\sin{\left(\gG_n\cdot\rr \right)},
\end{split}
\end{equation}
for all conduction-band parameters, and a second-order expansion for the valence band parameters. The latter is necessary to correctly reproduce the effective gauge (moir\'e) potential for valence-band electrons defined by the effective model $H_{v}(\rr)$. A detailed discussion of this can be found in Appendix \ref{app:landscapes}.

\begin{table}[h!]
\caption{Interlayer distance interpolation parameters entering Eq.\ \eqref{eq:d_interpolation}.}
\begin{center}
\begin{tabular}{P{1.0cm}  Q{1.5cm}@{.}R{1.5cm}  Q{1.5cm}@{.}R{1.5cm}}
\hline\hline
$n$ & \multicolumn{2}{P{3.0cm}}{$d_n^s\,[\mathrm{\AA}]$} & \multicolumn{2}{P{3.0cm}}{$d_n^a\,[\mathrm{\AA}]$}\\
\hline\hline
1 & $-0$&$016$ & $-0$&$072$ \\
2 & $-0$&$124$ & $-0$&$150$ \\
3 & $0$&$088$ & $-0$&$021$ \\
4 & $0$&$088$ & $-0$&$062$ \\
\hline
\multicolumn{5}{P{7.0cm}}{$d_0 = 3.490\,{\rm \AA}$} \\
\hline\hline
\end{tabular}
\end{center}
\label{tab:d_fit}
\end{table}%!!!!!!!!!!!!!!!!

The interpolated band Hamiltonians $H_\alpha(\rr)$ now explicitly contain the superlattice periodicity through the moir\'e Bragg vectors $\gG_n$, and can be numerically diagonalized in a zone-folding scheme: Let $\kk$ be a vector of the mBZ shown in Fig.\ \ref{fig:CoverFigure}, and consider how $H_\alpha(\rr)$ acts on the monolayer plane-wave states ($\gG_{mn}\equiv m\gG_1+n\gG_2$)
\begin{equation}\label{eq:PlaneWaves}
\begin{split}
    \braketD{\rr}{\phi_{mn}^{\kk}}_b =& \frac{1}{\sqrt{\mathcal{S}}}
    \begin{pmatrix}
    e^{i(\kk + \gG_{mn})\cdot \rr} \\ 0
    \end{pmatrix},\\
    \braketD{\rr}{\phi_{mn}^{\kk}}_t =& \frac{1}{\sqrt{\mathcal{S}}}
    \begin{pmatrix}
    0 \\ e^{i(\kk + \gG_{mn})\cdot \rr}
    \end{pmatrix},
\end{split}
\end{equation}
where $\mathcal{S}$ are the monolayer surface areas, assumed equal. Taking the interlayer tunneling term as an example, we obtain the following matrix elements:
\begin{widetext}
\begin{equation}\label{InterME}
\begin{split}
    {}_{b}\langle \phi_{ij}^{\kk}|T_{\alpha}|\phi_{rs}^{\kk} \rangle_{t} =& \sum_{n=0}^{6} \int\frac{d^{2}r}{\mathcal{S}} \, t_{\alpha}^{(n)}e^{-q_{n}(d(\mathbf{r})-d_{0})}\cos{(\gG_{n}\cdot\rr)}e^{i(\gG_{rs}-\gG_{ij})\cdot\rr}\\
    \approx& \sum_{n=0}^{6}\frac{t_{\alpha}^{(n)}}{2}\Bigg[ \delta_{\gG_{ij},\gG_{rs}+\gG_{n}} + \delta_{\gG_{ij},\gG_{rs}-\gG_{n}}- \frac{q_{n}}{2}\sum_{m=1}^{4}(d_{m}^{s}+id_{m}^{a})(\delta_{\gG_{ij},\gG_{rs}+\gG_{n}+\gG_{m}}+\delta_{\gG_{ij},\gG_{rs}-\gG_{n}+\gG_{m}})\\
    & \qquad + \frac{q_{n}}{2}\sum_{m=1}^{4}(-d_{m}^{s}+id_{m}^{a})(\delta_{\gG_{ij},\gG_{rs}+\gG_{n}-\gG_{m}}+\delta_{\gG_{ij},\gG_{rs}-\gG_{n}-\gG_{m}})\\
    & \qquad+ \frac{q_{n}^{2}}{8}\sum_{m=1}^{4}\sum_{\ell=1}^{4}\Big\{ (d_{m}^{s}d_{\ell}^{s} + 2id_{m}^{s}d_{\ell}^{a} - d_{m}^{a}d_{\ell}^{a})(\delta_{\gG_{ij},\gG_{rs}+\gG_{n}+\gG_{m}+\gG_{\ell}} + \delta_{\gG_{ij},\gG_{rs}-\gG_{n}+\gG_{m}+\gG_{\ell}})\\
    & \qquad + (d_{m}^{s}d_{\ell}^{s} - 2id_{m}^{s}d_{\ell}^{a} + d_{m}^{a}d_{\ell}^{a})(\delta_{\gG_{ij},\gG_{rs}+\gG_{n}+\gG_{m}-\gG_{\ell}} + \delta_{\gG_{ij},\gG_{rs}-\gG_{n}+\gG_{m}-\gG_{\ell}})\\
    &\qquad + (d_{m}^{s}d_{\ell}^{s} + 2id_{m}^{s}d_{\ell}^{a} + d_{m}^{a}d_{\ell}^{a})(\delta_{\gG_{ij},\gG_{rs}+\gG_{n}-\gG_{m}+\gG_{\ell}} + \delta_{\gG_{ij},\gG_{rs}-\gG_{n}-\gG_{m}+\gG_{\ell}})\\
    &\qquad +(d_{m}^{s}d_{\ell}^{s} - 2id_{m}^{s}d_{\ell}^{a} - d_{m}^{a}d_{\ell}^{a})(\delta_{\gG_{ij},\gG_{rs}+\gG_{n}-\gG_{m}-\gG_{\ell}} + \delta_{\gG_{ij},\gG_{rs}-\gG_{n}-\gG_{m}-\gG_{\ell}})\Big\}\Bigg].
\end{split}
\end{equation}
\end{widetext}
    
\begin{figure*}[p!]
    \centering
    \includegraphics[width=1.0\textwidth]{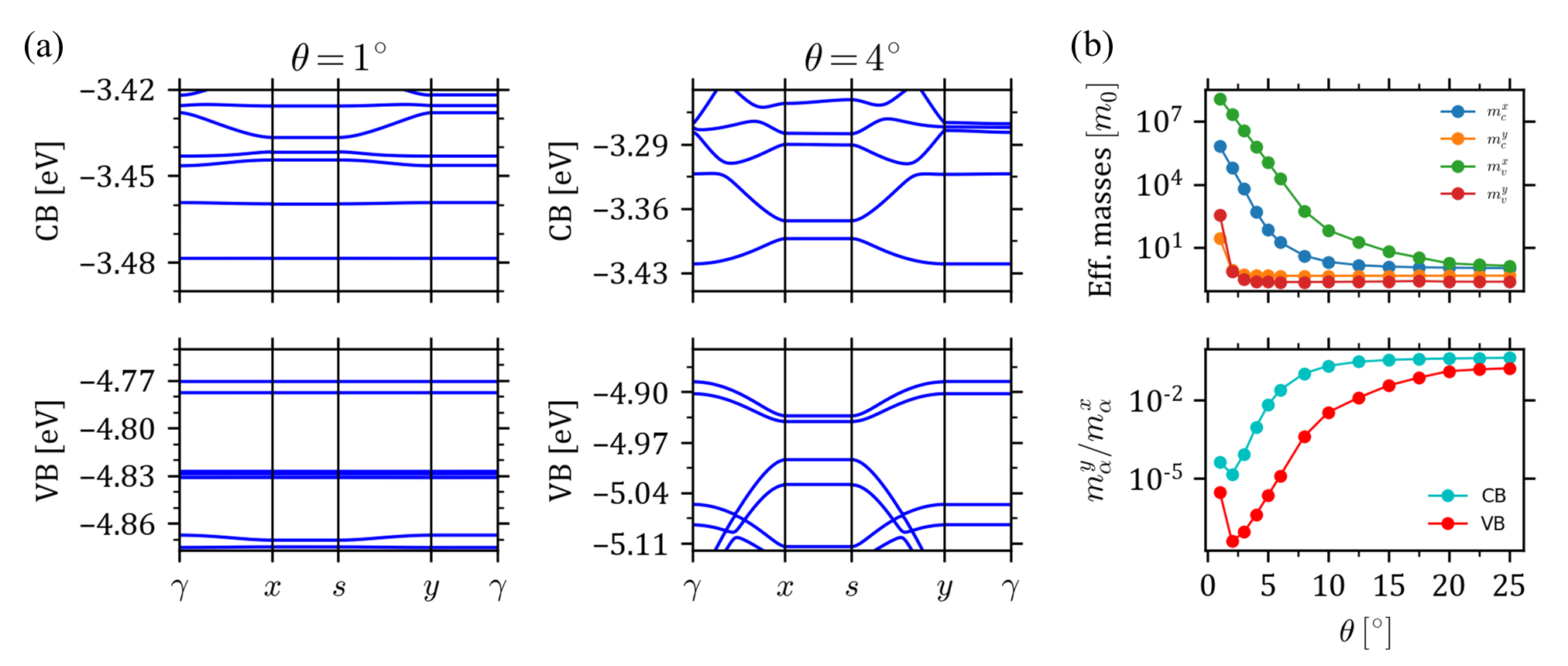}
    \caption{(a) Moir\'e miniband structure for twisted phosphorene bilayers with twist angles $\theta = 1^{\circ}$ (left panel) and $\theta=4^{\circ}$ (right panel), obtained by direct diagonalization of the continuous model Hamiltonian using the zone-folding approach. (b) Effective masses of the lowest conduction- and highest valence minibands around the $\gamma$-point (top panel), and mass ratio $m_\alpha^y/m_\alpha^x$, as functions of the twist angle.}
    \label{fig:Minibands}
\end{figure*}
\begin{figure*}[p!]
    \centering
    \includegraphics[width=1.0\textwidth]{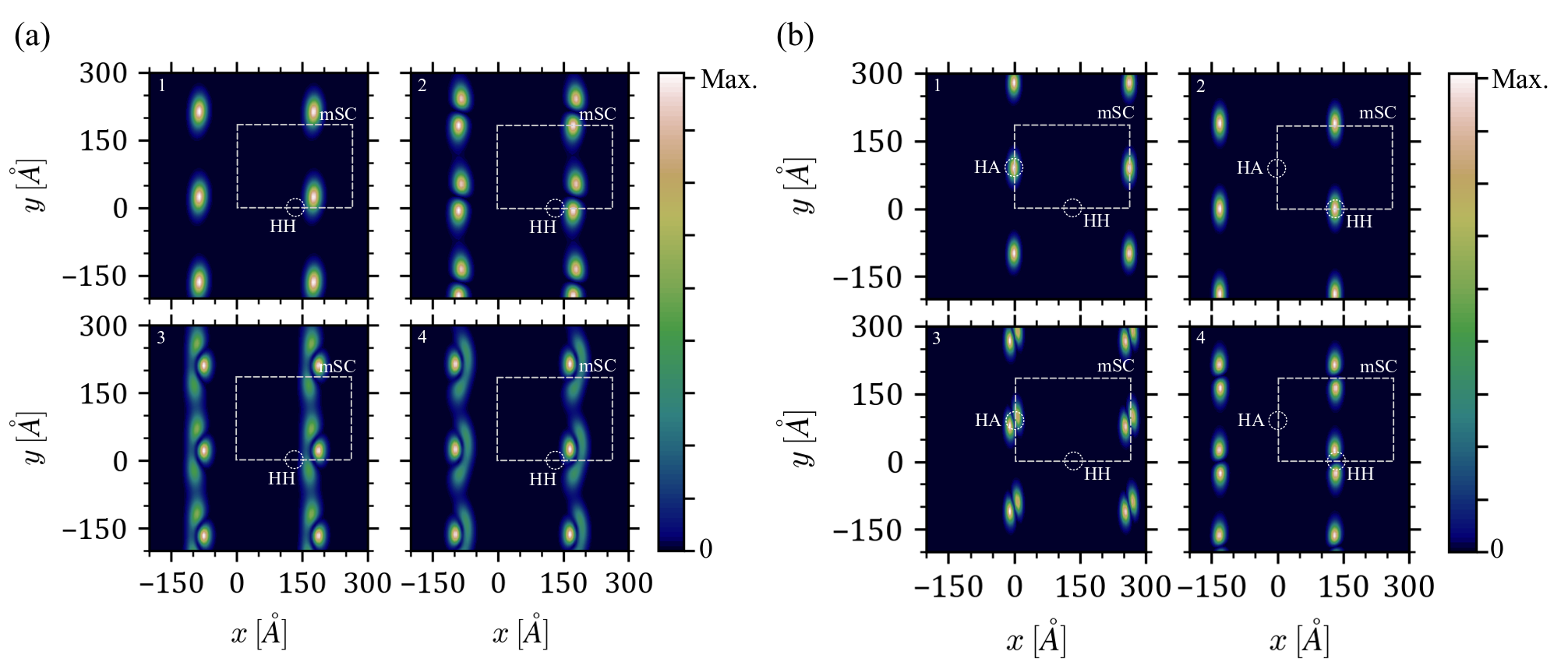}
    \caption{Spatial distribution of the first four (a) conduction- and (b) valence miniband eigenstates for $\theta=1^\circ$, computed as the moduli squared of the corresponding wave functions, averaged across the mBZ. Conduction minibands 1 and 2 correspond to arrays of quantum-dot-like states localized at HH regions of the mSC, whereas minibands 3 and 4 appear delocalized along the $\hat{\mathbf{y}}$ axis. All four valence minibands correspond to arrays of localized states alternating between two mSC locations: 1 and 3 appear at HA sites, whereas 2 and 4 appear at HH sites.}
    \label{fig:WFunctions}
\end{figure*}

Note that these matrix elements either conserve the total wave vector by setting $\gG_{ij}=\gG_{rs}$, or else couple states with wave vectors differing by a first- to sixth-star moir\'e Bragg vector, as $\gG_{ij}=\gG_{rs}+\gG_n$. Similarly, for the intralayer matrix elements we have [see Eq.\ \eqref{eq:Ediag}]
\begin{equation}\label{IntraME}
\begin{split}
    {}_{\lambda}\braket{\phi_{ij}^{\kk}|\varepsilon_{\alpha}^{\lambda}|\phi_{rs}^{\kk}}_{\lambda} =& \,\varepsilon_{\alpha}^{\lambda\,(0)}(\kk+\gG_{ij})\delta_{\gG_{ij},\gG_{rs}}\\
    &+ {}_{\lambda}\braket{\phi_{ij}^{\kk}|\delta\varepsilon_{\alpha}^{\lambda}|\phi_{rs}^{\kk}}_{\lambda},
\end{split}
\end{equation}
where the first term conserves the wave vector, and the matrix elements ${}_{\lambda}\braket{\phi_{ij}^{\kk}|\delta\varepsilon_{\alpha}^{\lambda}|\phi_{rs}^{\kk}}_{\lambda}$ have as similar structure to that of Eq.\ \eqref{InterME}. Note that we have generalized the Bloch function energy $\varepsilon_\alpha^{(0)}=\varepsilon_\alpha^{\rm ML}+\delta\varepsilon_\alpha^{(0)}$ to wave vectors close to the $\Gamma$ point as $\varepsilon_\alpha^{(0)} \rightarrow \varepsilon_\alpha^{t/b,(0)}(\kk+\gG_{mn})$, by introducing the kinetic energy terms
\begin{equation}
\begin{split}
    \varepsilon_\alpha^{t/b,(0)}(\qq) \equiv \varepsilon_\alpha^{\rm ML} + \delta\varepsilon_\alpha^{(0)} &+ \frac{\hbar^{2}|(\mathcal{R}_{\mp\theta/2}\qq)\cdot\hat{\mathbf{x}}|^{2}}{2m_{\alpha,x}^{(0)}}\\ &+ \frac{\hbar^{2}|(\mathcal{R}_{\mp\theta/2}\qq)\cdot\hat{\mathbf{y}}|^{2}}{2m_{\alpha,y}^{(0)}}.
\end{split}
\end{equation}
Here, $m_{c,x}^{(0)}=1.12m_{0}$, $m_{c,y}^{(0)}=0.46m_{0}$, $m_{v,x}^{(0)}=1.61m_{0}$ and $m_{v,y}^{(0)}=0.23m_{0}$ are the monolayer-band effective masses\cite{PhysRevLett.115.066403}, and the relative interlayer twist angle is included through the passive rotation of the monolayer dispersions.

 The zone-folding scheme consists in mapping all states \eqref{eq:PlaneWaves} lying outside the mBZ into minibands inside the mBZ as
 \begin{equation}
     \varepsilon_{mn}^{t/b}(\kk) \equiv \varepsilon_\alpha^{t/b,(0)}(\kk+\gG_{mn}),\quad \kk \in {\rm mBZ},
 \end{equation}
 labeled by the moir\'e vector indices $(m,n)$. The minibands couple vertically---\emph{i.e.}, conserving the mBZ wave vector $\kk$---amongst themselves according to the matrix elements \eqref{InterME} and \eqref{IntraME}, thus defining an eigenvalue problem that can be solved numerically for a finite number of minibands. In practice, we have found that the lowest conduction minibands are well converged for a range of indices $-10\le m,n \le 10$, corresponding to 882 basis states, for twist angles as small as $\theta=1^\circ$; whereas for the valence bands a total of $1250$ basis states ($-12\le m,n \le 12$) were needed. The resulting lowest (highest) energy eigenvalues represent the electronic energy spectra for the conduction (valence) bands around the $\gamma$ point of the mBZ.

\begin{figure}[t!]
    \centering
    \includegraphics[width=1.0\columnwidth]{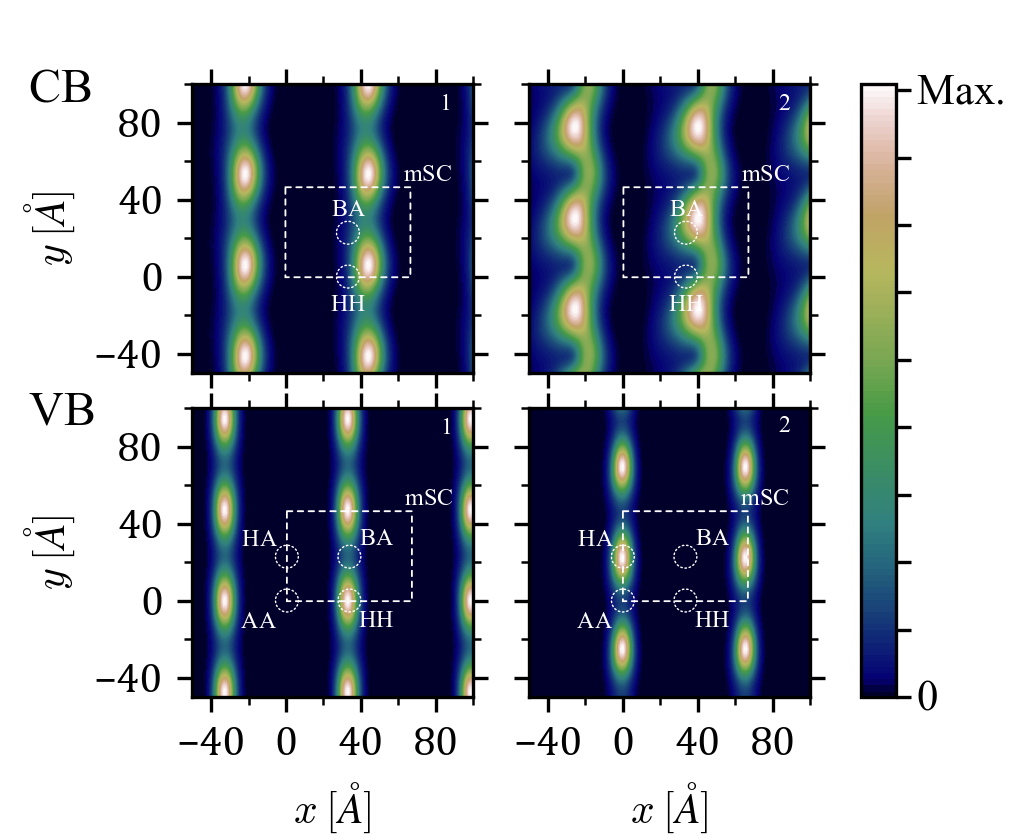}
    \caption{Spatial distribution of the first two conduction- (top) and valence (bottom) miniband eigenstates for $\theta=4^\circ$, computed as the moduli squared of the corresponding wave functions, averaged across the mBZ. All four minibands correspond to arrays of quasi-1D states that propagate along the $\hat{\mathbf{y}}$ direction.}
    \label{fig:WFunctions14}
\end{figure}

\section{Moir\'e minibands in twisted phosphorene bilayers}\label{sec:minibands}
Figure \ref{fig:Minibands}(a) shows the first few conduction- and valence minibands for twist angles $\theta=1^{\circ}$ and $4^{\circ}$. Flat minibands, corresponding to carrier states with vanishing group velocities, form for $\theta = 1^{\circ}$. To visualize the spatial distribution of the flat-miniband states, Fig.\ \ref{fig:WFunctions} shows their moduli squared averaged across the mBZ. The flat minibands correspond to arrays of spin-degenerate localized states, one per mSC, with the periodicity of the moir\'e superlattice. Conduction electrons localize near areas with HH stacking, whereas valence holes localize at ${\rm HA}$ regions of the mSC. In both cases, the localized states stretch along the $\hat{\mathbf{y}}$ axis, following the anisotropies of the corresponding moir\'e potentials, shown in Fig.\ \ref{fig:EnergyLand} of Appendix \ref{app:landscapes}. Thus, the localized wavefunctions approximately inherit the symmetry of the monolayer crystals, a feature that can be identified experimentally by scanning tunneling microscopy. This is most clearly observed in the lowest-energy conduction and highest-energy valence wavefunctions, which are $s$-like states stretched along the phosphorene unit cell's long axis (see Fig.\ \ref{fig:CoverFigure}). The second-lowest conduction miniband resembles a slightly rotated $p_y$ orbital, whereas the third and fourth ones clearly reflect the irregularities of the electron-confining potential, in particular its lack of an $\hat{\mathbf{x}} \rightarrow -\hat{\mathbf{x}}$ mirror symmetry. In the valence case, the second-highest miniband is also formed by $s$-like orbitals elongated in the $\hat{\mathbf{y}}$ axis, but localized at mSC ${\rm HH}$ regions, where a local minimum appears in the moir\'e potential (Fig.\ \ref{fig:EnergyLand}). This alternation between localization at ${\rm HA}$ and ${\rm HH}$ areas continues for the next two valence minibands, which resemble $p$ orbitals deformed by the confining potential anisotropy.

\begin{figure}[h!]
    \centering
    \includegraphics[width=0.9\columnwidth]{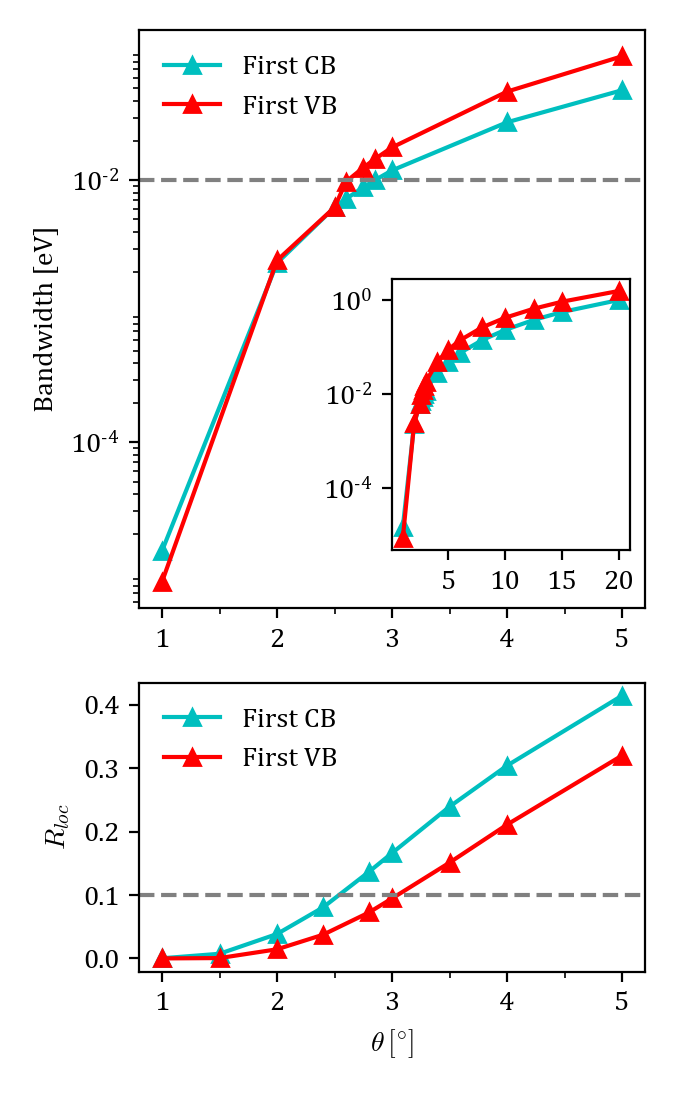}
    \caption{Bandwidths of the lowest conduction- and highest valence minibands (top), and delocalization coefficient along the $\hat{\boldsymbol{\mathrm{y}}}$ direction, $R_{\text{loc}}$ [see Eq.\ \eqref{eq:Rloc}] (bottom), as functions of the twist angle. The solid lines are guides to the eye.}
    \label{fig:BW-Deloc}
\end{figure}

When electron-electron interactions are taken into account, the localized conduction- and valence states described above for $\theta = 1^\circ$ suggest the realization of SU$(2)$ Hubbard model physics for both $\Gamma$-point electrons and holes, putting twisted phosphorene bilayers in the list of twistronic Hubbard materials, together with transition-metal dichalcogenide bilayer structures\cite{CorrelatedTMD1,CorrelatedTMD2}. We shall henceforth refer to this as the \emph{Hubbard regime}.

An altogether different regime is found for $\theta = 4^\circ$, where the lowest conduction- and highest valence minibands remain flat along the $\hat{\mathbf{x}}$ direction (along the segments $\overline{x\,s}$ and $\overline{y\,\gamma}$), but become dispersive in the $\hat{\mathbf{y}}$ direction---a consequence of the large mass anisotropies of the monolayer bands. Figure \ref{fig:WFunctions14} shows that these minibands correspond to arrays of quasi-one-dimensional states that propagate along the $\hat{\mathbf{y}}$ axis. For conduction electrons, these 1D states cross the HH and BA regions of the mSC, whereas valence holes exhibit a richer behavior. We find a nearly degenerate doublet of 1D states crossing the mSC at different regions: for the highest minibands, the 1D states propagate along the $\overline{{\rm HH}-{\rm BA}}$ segment, whereas the next highest miniband corresponds to a 1D state along the $\overline{{\rm AA}-{\rm HA}}$ line. In both cases, delocalization follows the shrinking of the mSC with increasing twist angle, which introduces a spatial overlap between neighboring localized states along the $\hat{\mathbf{y}}$ direction. We predict that these states should behave as coupled Tomonaga-Luttinger liquids\cite{Tomonaga,Luttinger} in the presence of electron-electron interactions, and we will refer to this as the \emph{Tomonaga-Luttinger regime}.

To establish the crossover between the Hubbard- an Tomonaga-Luttinger regimes, Fig.\ \ref{fig:BW-Deloc} shows two quantitative indicators of localization for the bottom conduction- and top valence minibands. In the top panel, we report their bandwidths for twist angles between $1^\circ$ and $5^\circ$, where a vanishing bandwidth is characteristic of the fully localized states of the Hubbard regime, whereas the Tomonaga-Luttinger regime exhibits a finite bandwidth. Based on typical experimental resolutions, we propose a 10 meV bandwidth as a reasonable value to define the crossover between the two regimes, which occurs around $\theta=2.85^\circ$ for the conduction band, and around $\theta=2.6^\circ$ for the valence band.

A consistent result is obtained by considering the probability density ratio
\begin{equation}\label{eq:Rloc}
    R_{\rm loc} = \overline{|\psi(\rr_{\rm inter})|^2}/\overline{|\psi(\rr_{\rm max})|^2},
\end{equation}
where $\overline{|\psi|^2}$ is the mBZ-averaged miniband wave-function squared reported in Figs.\ \ref{fig:WFunctions} and \ref{fig:WFunctions14}. The ratio \eqref{eq:Rloc} compares the state's weights at two different positions in an arbitrary mSC: $\rr_{\rm max}$, which is the localization site in the Hubbard regime (near HH for the conduction miniband, and HA for the valence miniband), and $\rr_{\rm inter}=\rr_{\rm max}+\mathbf{a}_1^{\rm M}/2$, corresponding to the position half way between that site and its replica in the next mSC along the positive $\hat{\mathbf{y}}$ axis. This ratio $R_{\rm loc}$ vanishes in the Hubbard limit, and tends to 1 deep in the Tomonaga-Luttinger limit. Setting the threshold between the two regimes at $R_{\rm loc}=0.1$, the bottom panel of Fig.\ \ref{fig:BW-Deloc} also establishes the crossover twist angles $\theta=2.4^\circ$ and $\theta=3^\circ$ for the conduction- and valence minibands, respectively.

Finally, our model predicts fully dispersive, anisotropic conduction- and valence minibands at twist angles $\theta \gtrsim 10^\circ$, which we call the \emph{ballistic regime}. However, as discussed in Sec.\ \ref{sec:model}, our continuous model represents a good approximation to the superlattice Hamiltonian only for small enough twist angles, for which the moir\'e periodicity is much larger than the atomic spacings. From Eq.\ \eqref{eq:aM} we can estimate that this condition roughly corresponds to $\theta < 8^\circ$, indicating that the crossover between the Tomonaga-Luttinger and ballistic regimes is not accurately described by our continuous model, including the specific crossover angles. Having forewarned the reader, we shall now assume that the continuous model can at least describe the ballistic regime at the qualitative level, and report our results. To quantify the anisotropies of the first conduction- and valence minibands, in Fig.\ \ref{fig:Minibands}(b) we plot their mass ratio $m_\alpha^{y}/m_\alpha^x$ for a wide range of twist angles. For the valence band, the mass ratio varies across three decades between $\theta = 10^\circ$ and $\theta=25^\circ$, indicating that the twist angle is an efficient knob for tuning the mass anisotropy of holes. A more modest but still quite significant variation of a factor of 10 is found for the mass ratio in the case of the lowest conduction miniband.

 The continuous superlattice model predictions summarized in this section are in excellent agreement with recent results based on large-scale DFT calculations, reported in Refs.\ \onlinecite{MoirePh2017} and \onlinecite{Brooks_2020}, with one exception: For the Tomonaga-Luttinger regime, both references identify the highest valence miniband with the 1D states propagating along the $\overline{{\rm AA}-{\rm HA}}$ segment of the mSC, which we identify as the second highest valence miniband. We attribute this quantitative discrepancy to the similar depths of the competing potential wells for holes appearing at HA and HH regions of the superlattice, as shown in Fig.\ \ref{fig:EnergyLand} of Appendix \ref{app:landscapes}. The energy order of the resulting states will necessarily depend on the fine quantitative details, which can differ in distinct DFT approximations.
 
 We remark that our numerical calculations, which rely on the model parametrization presented in Table \ref{tab:ModelParameters} and the zone-folding approach described in Sec.\ \ref{sec:interpolation}, were carried out at a low computational cost, and are easily reproducible. Moreover, the moir\'e gauge potential picture that derives from the continuum approximation to the superlattice Hamiltonian, discussed in detail in Appendix \ref{app:landscapes}, provides an intuitive picture for the origin of the low-dimensional states that emerge in the Hubbard and Tomonaga-Luttinger regimes. Admittedly, the validity of our continuous model in the large twist angle- or ballistic regime is questionable due to the reduced superlattice periodicity. However, the qualitative agreement between our results and the \emph{ab initio} calculations of Ref.\ \cite{Brooks_2020} lends some credibility to our extrapolation to large twist angles.

\section{Conclusions}\label{sec:conclusions}
In this paper, we have proposed a moir\'e superlattice Hamiltonian capable of describing the low-energy electronic spectra of twisted phosphorene bilayers. Numerical diagonalization of our model within a zone folding scheme has revealed three qualitatively distinct types of electronic states, depending on the twist angle: At small twist angles $\theta<2^\circ$, electrons and holes localize near regions of the mSC with approximate local HH- and HA stackings, respectively, giving mesoscale realizations of the SU(2) Hubbard model on a rectangular lattice, and motivating the term \emph{Hubbard regime}. At intermediate angles $2^\circ<\theta \lesssim 10^\circ$, we predict the formation of arrays of quasi-1D states that propagate across the mSC, along the long axis of the phosphorene unit cell. Each of these states can potentially exhibit Luttinger liquid properties, which motivates the term \emph{Tomonaga-Luttinger regime}. Finally, fully dispersive minibands are recovered at large twist angles $\theta \gtrsim 10^\circ$, which we call the \emph{ballistic regime}. In this case, we propose the twist angle as an efficient knob for tuning the miniband anisotropies, modulating their mass ratios by up to a factor of $10^1$ in the conduction case, and of $10^3$ in the valence case.  We believe that observation of these regimes is well within current experimental capabilities, by a combination of scanning tunneling microscopy and transport experiments. All of our results are in good agreement with large-scale \emph{ab initio} calculations found in the recent literature\cite{MoirePh2017,Brooks_2020,Wang_trilayerP}. The effective models developed in this paper are easily reproducible at a low computational cost, and motivate an intuitive understanding of the emergence of low-dimensional states based on the local properties of the different mSC regions.

To construct our superlattice Hamiltonian, we have formulated symmetry-based effective models for the lowest conduction- and highest valence $\Gamma$-point subband states of aligned phosphorene bilayers with arbitrary stacking. These models have been parametrized based on PBE + Grimme-D3 DFT calculations, supplemented with scissor corrections for the band gap, and fully taking into account out-of-plane relaxation of the bilayer system. We propose that these models may be used in their own right to describe the band structures of HA, HH and AA bilayer phosphorene domains. Analogous to the cases of twisted bilayer graphene\cite{Grelax1,Grelax2} and twisted transition-metal dichalcogenide homo- and heterobilayers\cite{TMDrelax1,TMDrelax2}, strong in-plane relaxation may also occur in twisted phosphorene bilayers at small enough twist angles, resulting in the formation of large HA domains, surrounded by smaller HH and AA ones, which can be individually described by the effective models discussed in this paper.

\acknowledgments{I.S.\ acknowledges financial support from CONACyT, through a \emph{Becas Nacionales} graduate scholarship. J.G-S. is thankful for the support provided by DGAPA-UNAM Project No.\ IA100822. F.M. acknowledges funding from DGAPA-UNAM through Grant Papiit No.\ IN113920. D.A.R-T. acknowledges funding from CONACyT Grants No.\ A1-S-14407 and No.\ 1564464. DFT calculations were performed at the DGCTIC-UNAM Supercomputing Center, through Project No.\ LANCAD-UNAM-DGCTIC-368. J.G-S.\ thanks A.\ Rodr\'iguez-Guerrero for the technical support provided throughout the project development. Finally, D.A.R-T. would like to thank V.\ I.\ Fal'ko for useful comments during the preparation of this paper.}

\bibliography{references}

\appendix

\section{DFT band structures for intermediate configurations}\label{app:indirectgap}
\begin{figure*}
    \centering
    \includegraphics{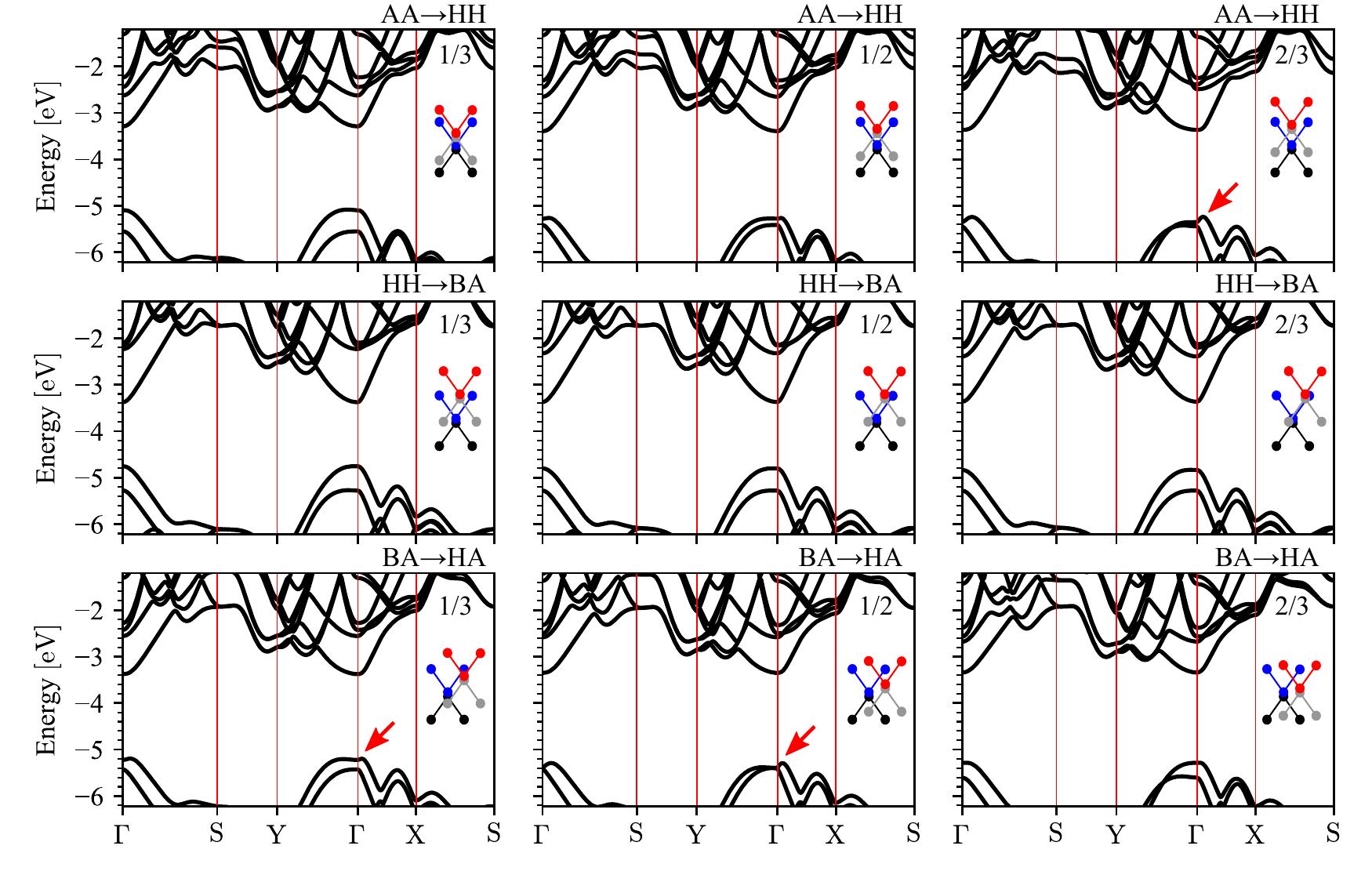}
    \caption{Scissor-corrected PBE + Grimme-D3 DFT band structures for stacking configurations intermediate to the high-symmetry cases listed in Table \ref{tab:r0}. In each case, the stacking configuration is labeled as the portion ($1/3$, $1/2$ or $2/3$) of the line segment between two high-symmetry configurations (\emph{e.g.}, ${\rm AA}\rightarrow {\rm HH}$) at which the corresponding value of $\rr_0$ is obtained, and the bilayer unit cell is sketched for clarity.}
    \label{fig:DFT3}
\end{figure*}
Figure \ref{fig:DFT3} shows the band structures for intermediate configurations to the high-symmetry stackings listed in Table \ref{tab:r0}, following the path ${\rm AA}\rightarrow {\rm HH} \rightarrow {\rm BA} \rightarrow {\rm HA}$ for $\rr_0$ values. These band structures, like those reported in Fig.\ \ref{fig:DFT2}, were obtained from PBE + Grimme-D3 DFT calculations, with a scissor-corrected band gap based on Ref.\ \cite{Castellanos_Gomez_2014}. By contrast to the high-symmetry stacking cases, we find five intermediate configurations where the band gap becomes slightly indirect, due to the appearance of a global valence band maximum close to the $\Gamma$ point, along the $\overline{\Gamma\,{\rm X}}$ line of the BZ. In three out of these five cases this maximum is significantly higher in energy than the $\Gamma$-point valence state, reaching a maximum energy difference of $\approx 115\,{\rm meV}$.

\section{Symmetry constraints for the Fourier components of the valence- and conduction band Bloch functions}\label{app:symmetry}
We begin by noting that [see Eqs.\ \eqref{eq:3stars}] $\GG_1\cdot\rr_{\mathcal{T}}=\GG_2\cdot\rr_{\mathcal{T}}=\pi$, such that
\begin{equation*}
    \GG_3\cdot\rr_{\mathcal{T}} = 0,\, \GG_4\cdot\rr_{\mathcal{T}} = \GG_5\cdot\rr_{\mathcal{T}} = \GG_6\cdot\rr_{\mathcal{T}} = 2\pi.
\end{equation*}
This immediately gives
\begin{equation}\label{eq:phases}
    e^{i\GG_n\cdot\rr_{\mathcal{T}}} = \left\{
    \begin{array}{rcl}
    -1 &,& n=1,\,2\\ \\
    1 &,& n=3,4,5,6
    \end{array}\right.,
\end{equation}
entering Eqs.\ \eqref{eq:ConstraintC_2z} and \eqref{eq:ConstraintC_2y}. Combining Eqs.\ \eqref{eq:ConstraintC_2z} and \eqref{eqs:Constraints} with Eq.\ \eqref{eq:phases} we obtain the following constraints for the Fourier coefficients $u_\alpha(\GG_n,z)$:
\begin{subequations}\label{eqs:all_constraints}
\begin{equation}
    u_{v}(\GG_1,z) = u_{v}(-\GG_1,z) = u_{v}(\GG_1,-z) ,
\end{equation}
\begin{equation}
    u_{v}(\GG_2,z) = u_{v}(-\GG_2,z) = u_{v}(\GG_2,-z),
\end{equation}
\begin{equation}
    u_{v}(\GG_3,z) = -u_{v}(-\GG_3,z) = -u_{v}(\GG_3,-z),
\end{equation}
\begin{equation}
    u_{v}(\GG_4,z) = -u_v(\GG_3,z),
\end{equation}
\begin{equation}
    u_{v}(\GG_6,z) = -u_{v}(-\GG_6,z) = u_{v}(\GG_6,-z),
\end{equation}
\begin{equation}
    u_c(\boldsymbol{0},z) = u_c(-\boldsymbol{0},z) = -u_c(\boldsymbol{0},-z),
\end{equation}
\begin{equation}
    u_{c}(\GG_2,z) = -u_{c}(-\GG_2,z) = u_{c}(\GG_2,-z),
\end{equation}
\begin{equation}
    u_{c}(\GG_3,z) = u_{c}(-\GG_3,z) = -u_{c}(\GG_3,-z),
\end{equation}
\begin{equation}
    u_{c}(\GG_4,z) = u_c(\GG_3,z),
\end{equation}
\begin{equation}
    u_{c}(\GG_5,z) = u_{c}(-\GG_5,z) = -u_{c}(\GG_5,-z),
\end{equation}
\begin{equation}
    u_{c}(\GG_6,z) = u_{c}(-\GG_6,z) = -u_{c}(\GG_6,-z).
\end{equation}
\begin{equation}
    u_{v}(\boldsymbol{0},z) = u_{v}(\pm\GG_5,z) = u_c(\pm\GG_1,z) = 0.
\end{equation}
\end{subequations}

\section{Matrix elements of the crystal potential}\label{app:crystalpotential}
As shown in Eq.\ \eqref{eq:Heffelems}, the intralayer matrix elements of the effective Hamiltonian $H_{\rm eff}$ for layer $\lambda$ correspond to those of the crystal potential of the opposite layer, $V_{\bar{\lambda}}$, given by
\begin{equation}
\begin{split}
    {}_{\lambda}\langle \alpha',\kk' |V_{\bar{\lambda}}&|\alpha,\kk \rangle_{\lambda} = \sum_{\GG,\GG',\GG''}\delta_{\kk-\kk',\GG'-\GG''-\GG}\\
    &\times  e^{i(\GG''-\GG-\kk+\GG'+\kk')\cdot\rr_{\lambda}}\\
    &\times \int dz\,u_{\alpha'}^\lambda(\GG',z)u_{\alpha}^\lambda(\GG,z)V_{\bar{\lambda}}(\GG'',z),
\end{split}
\end{equation}
where we have Fourier expanded the crystal potential as
\begin{equation}
    V_{\bar{\lambda}}(\rr-\rr_{\bar{\lambda}},z)=\sum_{\GG''}\frac{e^{i\GG'\cdot(\rr-\rr_{\bar{\lambda}})}}{\sqrt{N}}V_{\bar{\lambda}}(\GG,z),
\end{equation}
and used $\rr_{\bar{\lambda}} = - \rr_{\lambda}$. Once again, we consider only $\kk$ and $\kk'$ close to the $\Gamma$ point, and the momentum conservation condition simplifies to
\begin{equation*}
    \delta_{\kk',\kk}\delta_{\GG',\GG''+\GG}.
\end{equation*}
The only non vanishing matrix elements are then
\begin{equation}\label{eq:Velems}
\begin{split}
    &{}_{\lambda}\langle \alpha',\kk |V_{\bar{\lambda}}|\alpha,\kk \rangle_{\lambda} = \sum_{\GG,\GG',\GG''}\delta_{\GG',\GG''+\GG}e^{i(\GG''-\GG+\GG')\cdot\rr_{\lambda}}\\
    &\qquad \times \int dz\,  u_{\alpha'}^{\lambda*}(\GG',z)u_{\alpha}^\lambda(\GG,z)V_{\bar{\lambda}}(\GG'',z).
\end{split}
\end{equation}
At this point we approximate the Bloch functions and the potential by their first few Fourier components, taking only the first three stars of Bragg vectors. In addition, we shall focus on the case $\alpha'=\alpha$, since intralayer-interband transitions are strongly suppressed by the band gap. We split Eq.\ \eqref{eq:Velems} into the following contributions: When $\GG=\GG'=\GG''=\boldsymbol{0}$ or $\GG=\GG'\ne\boldsymbol{0}$ with $\GG''=\boldsymbol{0}$, we get
\begin{equation}\label{eq:V0}
\begin{split}
    [V_{\bar{\lambda}}^{(0)}]_{\alpha,\kk} =& \int dz\,  u_{\alpha}^{\lambda *}(\boldsymbol{0},z)u_{\alpha}^\lambda(\boldsymbol{0},z)V_{\bar{\lambda}}(\boldsymbol{0},z)\\
    &+\sum_{|n|=1}^6\int dz\,u_{\alpha}^{\lambda*}(\GG_n,z)u_{\alpha}^\lambda(\GG_n,z)V_{\bar{\lambda}}(\boldsymbol{0},z),
\end{split}
\end{equation}
which is independent of stacking. Then, we consider $\GG=\boldsymbol{0}$ with $\GG''=\GG'\ne\boldsymbol{0}$, and $\GG'=0$ with $\GG''=-\GG\ne\boldsymbol{0}$, which give
\begin{equation}\label{eq:V1}
\begin{split}
    [V_{\bar{\lambda}}^{(1)}]_{\alpha,\kk} = 2\sum_{|n|=1}^6\int dz\,&\mathrm{Re}\left\{  u_{\alpha}^{\lambda *}(\GG_n,z)u_{\alpha}^\lambda(\boldsymbol{0},z) \right.\\
    &\times \left. e^{2i\GG_n\cdot \rr_{\lambda}} \right\}\,V_{\bar{\lambda}}(\GG_n,z),
\end{split}
\end{equation}
where using the conditions $V_{\bar{\lambda}}(-\GG,z)=V_{\bar{\lambda}}^*(\GG,z)$, required for a real-valued crystal potential, and $C_{2\hat{\mathbf{z}}}V_{\bar{\lambda}}(\GG,z)=V_{\bar{\lambda}}(-\GG,z)=V_{\bar{\lambda}}(\GG,z)$, required by rectangular symmetry, we have concluded that $V_{\bar{\lambda}}(\GG,z)\in\mathbb{R}$. Simplifying \eqref{eq:V1} and using the symmetry constraints \eqref{eqs:all_constraints}, we obtain for the conduction subbands
\begin{widetext}
\begin{equation}\label{eq:V1c}
\begin{split}
    [V_{\bar{\lambda}}^{(1)}]_{c,\kk} =& 4\int dz \Big\{\left[-\mathrm{Re} u_{c}^\lambda(\GG_2,z)\mathrm{Im}u_{c}^{\lambda}(\boldsymbol{0},z) + \mathrm{Im} u_{c}^\lambda(\GG_2,z)\mathrm{Re}u_{c}^{\lambda}(\boldsymbol{0},z) \right]V_{\bar{\lambda}}(\GG_2,z)\sin{\left(2\GG_2\cdot\rr_\lambda \right)}\\
    &+\left[\mathrm{Re} u_{c}^\lambda(\GG_3,z)\mathrm{Re}u_{c}^{\lambda}(\boldsymbol{0},z) + \mathrm{Im} u_{c}^\lambda(\GG_3,z)\mathrm{Im}u_{c}^{\lambda}(\boldsymbol{0},z) \right]V_{\bar{\lambda}}(\GG_3,z)\left[\cos{\left(2\GG_3\cdot\rr_\lambda \right)} + \cos{\left(2\GG_4\cdot\rr_\lambda \right)} \right]\\
    &+\left[\mathrm{Re} u_{c}^\lambda(\GG_5,z)\mathrm{Re}u_{c}^{\lambda}(\boldsymbol{0},z) + \mathrm{Im} u_{c}^\lambda(\GG_5,z)\mathrm{Im}u_{c}^{\lambda}(\boldsymbol{0},z) \right]V_{\bar{\lambda}}(\GG_5,z)\cos{\left(2\GG_5\cdot\rr_\lambda \right)}\\
    &+\left[\mathrm{Re} u_{c}^\lambda(\GG_6,z)\mathrm{Re}u_{c}^{\lambda}(\boldsymbol{0},z) + \mathrm{Im} u_{c}^\lambda(\GG_6,z)\mathrm{Im}u_{c}^{\lambda}(\boldsymbol{0},z) \right]V_{\bar{\lambda}}(\GG_6,z)\cos{\left(2\GG_6\cdot\rr_\lambda \right)} \Big\},
\end{split}
\end{equation}
having made use of the symmetry property of the potential $\sigma_{zx}V_{\bar{\lambda}}(\GG,z)=V_{\bar{\lambda}}(\GG,z)$, which gives $V_{\bar{\lambda}}(\GG_4,z)=V_{\bar{\lambda}}(\GG_3,z)$. By contrast, for the valence subbands this contribution vanished identically, due to the fact that $u_{v}(\boldsymbol{0},z)=0$.

Finally, we consider the case when $\GG'=\GG''+\GG$ with $\GG,\GG',\GG''\ne\boldsymbol{0}$, which gives
\begin{equation}\label{eq:V2}
\begin{split}
    [V_{\bar{\lambda}}^{(2)}]_{\alpha,\kk} =& \sum_{n,n'=1}^6\int dz\,  u_{\alpha}^{\lambda *}(\GG_{n'},z)u_{\alpha}^\lambda(\GG_n,z) \\
    &\times  V_{\bar{\lambda}}(\GG_{n'}-\GG_n,z) e^{2i(\GG_{n'}-\GG_n)\cdot \rr_{\lambda}}.
\end{split}
\end{equation}
In this last case, we must restrict ourselves to combinations $(n,n')$ such that $\GG_{n'}-\GG_n$ belongs to one of the first three stars of Bragg vectors, since we have assumed that $V_{\bar{\lambda}}(\GG_{n'}-\GG_n,z)$ is negligible otherwise. Every combination $(n,n)$ meets this requirement, giving a contribution which is independent of stacking, and can thus be grouped together with $[V_{\bar{\lambda}}^{(0)}]_{\alpha,\kk}$. The remaining possible combinations are listed in Table \ref{tab:nnp}.
\begin{table}[h!]
\caption{Combinations $(n',\,n)$ contributing to Eq.\ \eqref{eq:V2}.}
\begin{center}
\begin{tabular}{P{4.0cm} P{4.0cm} | P{4.0cm} P{4.0cm}}
\hline\hline
\, & $(n',\,n)$ & \, & $(n',\,n)$\\
 \hline\hline
 \, & $(4,2),\,(3,-2),\,(5,1),$ &  & $(-4,-2),\,(-3,2),\,(-5,-1),$\\
 \raisebox{1.5ex}[0pt]{$\GG_{n'}-\GG_n=\GG_1$} & $(2,-3),(-2,-4),(-1,-5)$ & \raisebox{1.5ex}[0pt]{$\GG_{n'}-\GG_n=-\GG_1$} & $(-2,3),(2,4),(1,5)$ \\
 \hline
 \, & $(1,3),\,(-1,-4),\,(4,1),$ &  & $(-1,-3),\,(1,4),\,(-4,-1),$\\
 \raisebox{1.5ex}[0pt]{$\GG_{n'}-\GG_n=\GG_2$} & $(-3,-1),(6,2),(-2,-6)$ & \raisebox{1.5ex}[0pt]{$\GG_{n'}-\GG_n=-\GG_2$} & $(3,1),(-6,-2),(2,6)$ \\
 
  \hline
 \, & $(1,2),\,(-2,-1),\,(5,4),$ &  & $(-1,-2),\,(2,1),\,(-5,-4)$\\
 \raisebox{1.5ex}[0pt]{$\GG_{n'}-\GG_n=\GG_3$} & $(-6,-4),\,(4,6),\,(-4,-5)$ & \raisebox{1.5ex}[0pt]{$\GG_{n'}-\GG_n=-\GG_3$} & $(6,4),\,(-4,-6),\,(4,5)$ \\
 
 \hline
 \, & $(1,-2),\,(2,-1),\,(5,3),$ &  & $(-1,2),\,(-2,1),\,(-5,-3),$\\
 \raisebox{1.5ex}[0pt]{$\GG_{n'}-\GG_n=\GG_4$} & $(3,-6),\,(6,-3),\,(-3,-5)$ & \raisebox{1.5ex}[0pt]{$\GG_{n'}-\GG_n=-\GG_4$} & $(-3,6),\,(-6,3),\,(3,5)$ \\
 
 \hline
 $\GG_{n'}-\GG_n=\GG_5$ & $(1,-1),\,(4,-3),\,(3,-4)$ & $\GG_{n'}-\GG_n=-\GG_5$ & $(1,-1),\,(4,-3),\,(3,-4)$\\
 
 \hline
 $\GG_{n'}-\GG_n=\GG_6$ & $(2,-2),\,(4,3),\,(-3,-4)$ & $\GG_{n'}-\GG_n=-\GG_6$ & $(-2,2),\,(-4,-3),\,(3,4)$\\
 
\hline\hline
\end{tabular}
\end{center}
\label{tab:nnp}
\end{table}%!!!!!!!!!!!!!!!!

Following Table \ref{tab:nnp} and Eq.\ \eqref{eqs:all_constraints}, we obtain for the conduction subbands
\begin{equation}\label{eq:V2c}
\begin{split}
        [V_{\bar{\lambda}}^{(2)}]_{c,\kk}=&\int dz\,\Big\{-4\mathrm{Im}\left[u_c^{\lambda *}(\GG_6,z)u_c^\lambda(\GG_2,z)\right]V_{\bar{\lambda}}(\GG_2)\sin{\left(2\GG_2\cdot\rr_\lambda \right)}\\
        &+4\mathrm{Re}\left[u_c^{\lambda *}(\GG_6,z)u_c^\lambda(\GG_3,z)+u_c^{\lambda *}(\GG_5)u_c^\lambda(\GG_3)\right]V_{\bar{\lambda}}(\GG_3)\left[\cos{\left(2\GG_3\cdot\rr_\lambda \right)}+\cos{\left(2\GG_4\cdot\rr_\lambda \right)}\right]\\
        &+4|u_c^{\lambda}(\GG_3,z)|^2V_{\bar{\lambda}}(\GG_5)\cos{\left(2\GG_5\cdot\rr_\lambda \right)}+2\left[2|u_c^{\lambda}(\GG_3,z)|^2-|u_c^\lambda(\GG_2,z)|^2\right]V_{\bar{\lambda}}(\GG_6)\cos{\left(2\GG_6\cdot\rr_\lambda \right)}\Big\},
\end{split}
\end{equation}
whereas for the valence subbands we get
\begin{equation}\label{eq:V2v}
\begin{split}
        [V_{\bar{\lambda}}^{(2)}]_{v,\kk}=&\int dz\,\Big\{-4\left(2\mathrm{Im}\left[u_v^{\lambda *}(\GG_1,z)u_v^\lambda(\GG_3,z)\right]+\mathrm{Im}\left[u_v^{\lambda *}(\GG_6,z)u_v^\lambda(\GG_2,z) \right]\right)V_{\bar{\lambda}}(\GG_2,z)\sin{\left(2\GG_2\cdot\rr_\lambda \right)}\\
        &+2\left(\mathrm{Re}\left[u_v^{\lambda *}(\GG_1,z)u_v^\lambda(\GG_2,z) \right] - \mathrm{Re}\left[u_v^{\lambda *}(\GG_6,z)u_v^\lambda(\GG_3,z) \right] \right)V_{\bar{\lambda}}(\GG_3,z)\left[\cos{\left(2\GG_3\cdot\rr_\lambda \right)}+\cos{\left(2\GG_4\cdot\rr_\lambda \right)} \right]\\
        &+2\left(|u_v^{\lambda *}(\GG_1,z)|^2 + 2|u_v^{\lambda *}(\GG_3,z)|^2 \right)V_{\bar{\lambda}}(\GG_5,z)\cos{\left(2\GG_5\cdot\rr_\lambda \right)}\\
        &+2\left(|u_v^{\lambda *}(\GG_2,z)|^2 - 2|u_v^{\lambda *}(\GG_3,z)|^2 \right)V_{\bar{\lambda}}(\GG_6,z)\cos{\left(2\GG_6\cdot\rr_\lambda \right)}\Big\}.
\end{split}
\end{equation}
Equations \eqref{eq:V0}, \eqref{eq:V1c}, \eqref{eq:V2c} and \eqref{eq:V2v} may be summarized as
\begin{equation}\label{eq:AppV}
\begin{split}
    {}_{\lambda}\langle \alpha,\kk |V_{\bar{\lambda}}|\alpha,\kk \rangle_{\lambda} =& v_\alpha^{(0)} + v_\alpha^{(2)}\sin{\left(2\GG_2\cdot\rr_\lambda \right)} + \sum_{n=3}^6 v_\alpha^{(n)} \cos{\left(2\GG_n\cdot\rr_{\lambda} \right)},
\end{split}
\end{equation}
with $v_\alpha^{(4)}=v_\alpha^{(3)}$.
\end{widetext}

\section{Virtual tunneling corrections and L\"owdin partitioning}\label{app:virtual}
Here, we shall consider tunneling matrix elements between the Bloch state $|\alpha,\kk \rangle_\lambda$ in layer $\lambda$, and $|\beta,\kk \rangle_{\bar{\lambda}}$ in the opposite layer. For $\alpha=v$, we consider only $\beta=c$, which gives
\begin{equation}\label{eq:ExtraHyb_v}
\begin{split}
    &T_{c,v}^*(\rr_0) \equiv {}_t\langle c,\kk | H_{\rm eff} |v,\kk \rangle_b = \sum_{n=2}^6t_{c,v}^{(n)}\sin{\left(\GG_n\cdot\rr_0 \right)},\\
    &T_{v,c}^*(\rr_0) \equiv {}_t\langle v,\kk | H_{\rm eff} |c,\kk \rangle_b = - T_{c,v}^*(\rr_0),\\
    &t_{c,v}^{(5)} = 0,\,t_{c,v}^{(4)}=-t_{c,v}^{(3)}.
\end{split}
\end{equation}
For $\alpha=c$, we consider $\beta=v,c'$, and obtain
\begin{equation}\label{eq:ExtraHyb_c}
\begin{split}
    &T_{\beta,c}^*(\rr_0) \equiv {}_t\langle \beta,\kk | H_{\rm eff} |c,\kk \rangle_b = \sum_{n=2}^6t_{\beta,c}^{(n)}\sin{\left(\GG_n\cdot\rr_0 \right)},\\
    &T_{c,\beta}^*(\rr_0) \equiv {}_t\langle c,\kk | H_{\rm eff} |\beta,\kk \rangle_b = - T_{\beta,c}^*(\rr_0),\\
    &t_{\beta,c}^{(5)} = 0,\,t_{\beta,c}^{(4)}=-t_{\beta,c}^{(3)}.
\end{split}
\end{equation}
In both cases, we have the definition
\begin{equation}
    t_{\beta,\alpha}^{(n)}=2i\int dz\,u_\beta^{\bar{\lambda}*}(\GG_n,z)u_\alpha^\lambda(\GG_n,z).
\end{equation}
The matrix elements \eqref{eq:ExtraHyb_c} have the same form for either $\beta=v,c'$, as well as the same form as those of Eqs.\ \eqref{eq:ExtraHyb_v}, because all cases involve tunneling between a band that transforms like representation $B_{1u}$ of group $D_{2h}$ (band $c$), and another that transforms like representation $B_{3g}$ (bands $v$ and $c'$).

The total Hamiltonian, involving two layers with three bands each, takes the form
\begin{equation}\label{eq:H6by6}
    H_{\rm eff}^{6\times6}=\begin{pmatrix}
    \varepsilon_c^0 & T_c & 0 & T_{c,v} & 0 & T_{c,c'}\\
    T_c^* & \varepsilon_c^0 & -T_{c,v}^* & 0 & -T_{c,c'}^* & 0\\
    0 & -T_{c,v} & \varepsilon_v^0 & T_v & 0 & 0\\
    T_{c,v}^* & 0 & T_v^* & \varepsilon_v^0 & 0 & 0\\
    0 & -T_{c,c'} & 0 & 0 & \varepsilon_{c'}^0 & T_{c'}\\
    T_{c,c'}^* & 0 & 0 & 0 & T_{c'}^* & \varepsilon_{c'}^0
    \end{pmatrix},
\end{equation}
with the basis ordering $\{|c\rangle_b,|c\rangle_t,|v \rangle_b, | v\rangle_t,|c' \rangle_b,|c' \rangle_t\}$. In Eq.\ \eqref{eq:H6by6}, we have introduced the energy of the monolayer $\Gamma$-point state of band $c'$, $\varepsilon_{c'}^0$, and the tunneling matrix element $T_{c'}^*$, which has the same form as Eq.\ \eqref{eqs:MelemsVB}.

Next, we project out the block $\{|c'\rangle_b,\,|c'\rangle_t\}$ up to second order in perturbation theory using L\"owdin's partitioning\cite{WinklerBook}, resulting in
\begin{equation}
    H_{\rm eff}^{4\times4}=\begin{pmatrix}
    \varepsilon_c^0-\frac{|T_{c,c'}|^2}{\varepsilon_{c'}^0-\varepsilon_{c}^0} & T_c & 0 & T_{c,v} \\
    T_c^* & \varepsilon_c^0-\frac{|T_{c,c'}|^2}{\varepsilon_{c'}^0-\varepsilon_{c'}^0} & -T_{c,v}^* & 0\\
    0 & -T_{c,v} & \varepsilon_v^0 & T_v \\
    T_{c,v}^* & 0 & T_v^* & \varepsilon_v^0
    \end{pmatrix},
\end{equation}
and an independent $2\times2$ Hamiltonian for the $c'$ band sector, which is not of interest to us at the moment. Repeating the projection procedure, this time to eliminate the interlayer coupling between conduction and valence bands at second order in perturbation theory, we obtain
\begin{subequations}
\begin{equation}
    H_{\rm eff}^{c}\approx\begin{pmatrix}
    \varepsilon_c^0-\frac{|T_{c,c'}|^2}{\varepsilon_{c'}^0-\varepsilon_{c}^0} + \frac{|T_{c,v}|^2}{\varepsilon_c^0 - \varepsilon_v^0} & T_c \\
    T_c^* & \varepsilon_c^0-\frac{|T_{c,c'}|^2}{\varepsilon_{c'}^0-\varepsilon_{c'}^0}+ \frac{|T_{c,v}|^2}{\varepsilon_c^0 - \varepsilon_v^0}
    \end{pmatrix}
\end{equation}
\begin{equation}
    H_{\rm eff}^{v}\approx\begin{pmatrix}
    \varepsilon_v^0-\frac{|T_{c,v}|^2}{\varepsilon_{c}^0-\varepsilon_{v}^0} & T_v \\
    T_v^* & \varepsilon_v^0-\frac{|T_{v,c}|^2}{\varepsilon_{c}^0-\varepsilon_{v}^0}
    \end{pmatrix},
\end{equation}
\end{subequations}
where we have approximated
\begin{equation}
    \varepsilon_v^0 - \varepsilon_{c}^0 + \frac{|T_{c,c'}|^2}{\varepsilon_{c'}^0-\varepsilon_c^0} \approx \varepsilon_v^0 - \varepsilon_{c}^0.
\end{equation}
Expanding, \emph{e.g.}, $|T_{c,v}|^2$, we obtain
\begin{widetext}
\begin{equation}\label{eq:FullT2}
\begin{split}
    \frac{|T_{c,v}|^2}{4} =& \, \frac{|u_c^{\bar{\lambda}*}(\GG_2,z)u_v^{\lambda}(\GG_2,z)|^2+|u_c^{\bar{\lambda}*}(\GG_6,z)u_v^{\lambda}(\GG_6,z)|^2}{2}\\
    &+2\mathrm{Re}\left[u_c^{\bar{\lambda}*}(\GG_2,z)u_v^{\lambda}(\GG_2,z) u_c^{\bar{\lambda}}(\GG_3,z)u_v^{\lambda *}(\GG_3,z) \right]\cos{\left(\GG_1\cdot\rr_0 \right)}\\
    &+2\mathrm{Re}\left[u_c^{\bar{\lambda}*}(\GG_2,z)u_v^{\lambda}(\GG_2,z) u_c^{\bar{\lambda}}(\GG_6,z)u_v^{\lambda *}(\GG_6,z) \right]\cos{\left(\GG_2\cdot\rr_0 \right)}\\
    &-\mathrm{Re}\left[u_c^{\bar{\lambda}*}(\GG_3,z)u_v^{\lambda}(\GG_3,z) u_c^{\bar{\lambda}}(\GG_6,z)u_v^{\lambda *}(\GG_6,z) \right]\left[\cos{\left(\GG_3\cdot\rr_0 \right)} + \cos{\left(\GG_4\cdot\rr_0 \right)} \right]\\
    &+\frac{|u_c^{\bar{\lambda}*}(\GG_3,z)u_v^{\lambda}(\GG_3,z)|^2}{2}\cos{\left(\GG_6\cdot\rr_0 \right)},
\end{split}
\end{equation}
\end{widetext}
and $|T_{c,c'}|^2$ has the same form, simply exchanging $v\rightarrow c'$ everywhere in \eqref{eq:FullT2}.

These results can be summarized as
\begin{equation}\label{eq:LowdinAll}
    H_{\rm eff}^\alpha=\begin{pmatrix}
    \varepsilon_\alpha^0 + \delta \varepsilon_\alpha(\rr_0) & T_\alpha(\rr_0)\\
    T_\alpha^{*}(\rr_0) & \varepsilon_\alpha^0 + \delta \varepsilon_\alpha(\rr_0) 
    \end{pmatrix},
\end{equation}
for $\alpha=c,v$, with the definitions
\begin{equation}
    \delta \varepsilon_\alpha(\rr_0) =  w_\alpha^{(0)} + \sum_{n=1}^6 w_\alpha^{(n)}\cos{\left(\GG_n\cdot\rr_0\right)},
\end{equation}
and setting $w_\alpha^{(4)}=w_\alpha^{(3)}$ and $w_\alpha^{(5)}=0$. Comparing \eqref{eq:LowdinAll} with Eqs. \eqref{eq:Hform} and \eqref{eq:diagform} yields Eq.\ \eqref{eq:dE}.

\section{Effective moir\'e gauge potentials}\label{app:landscapes}
\begin{figure}[h!]
    \centering
    \includegraphics[width=0.8\columnwidth]{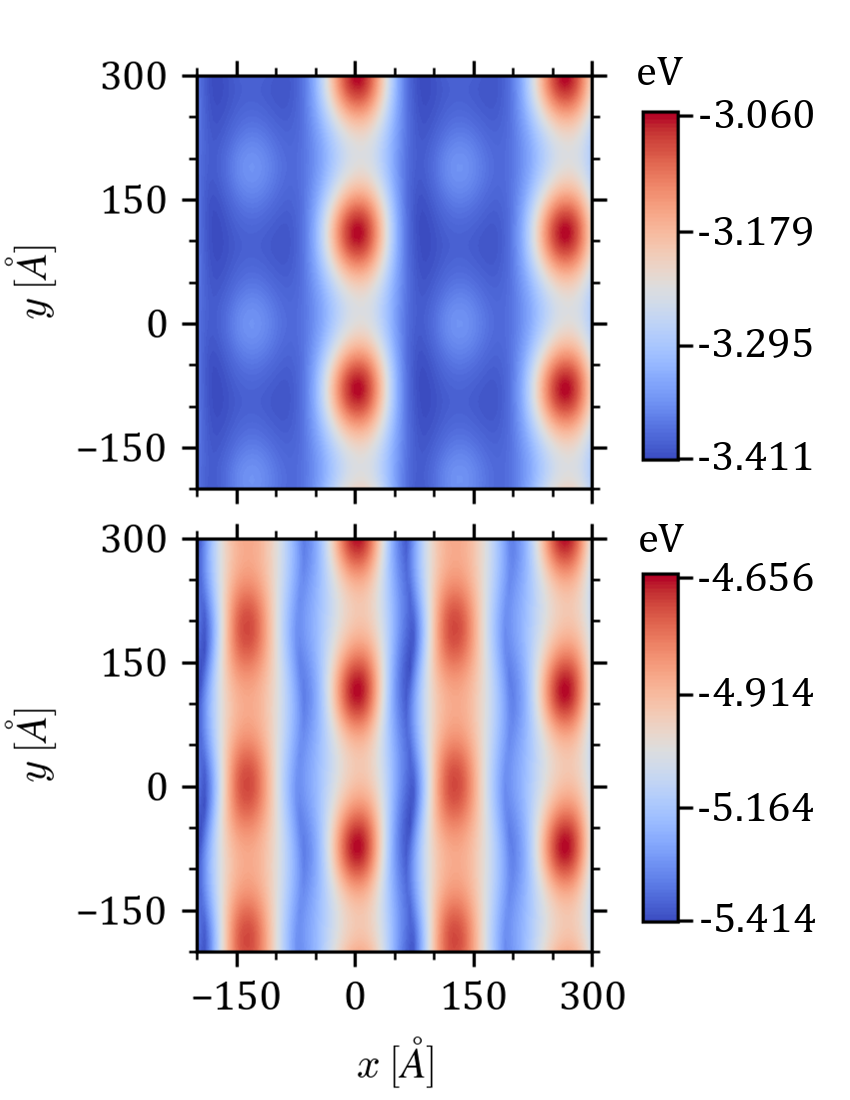}
    \caption{Effective moir\'e potential landscapes for conduction- (top) and valence-band (bottom) electrons. Conduction electrons become localized by the potential wells forming at HH regions, colored dark blue in the top panel. Conversely, valence-band holes are localized by the competing potential barriers shown in dark red in the middle panel, appearing at HA and HH superlattice regions.}
    \label{fig:EnergyLand}
\end{figure}
The effective Hamiltonians $H_\alpha(\rr)$, obtained by substituting $\rr_0=\theta\hat{\mathbf{z}}\times\rr$ into Eq.\ \eqref{eq:Hform}, constitute effective gauge potentials\cite{niu2017physical,LandscapesAPL} for the phosphorene electronic states. In the limit of an infinite superlattice periodicity, the motion of an $\alpha$-band electron can be described by $H_\alpha(\vv)$, where the position $\rr$ is replaced by a parameter $\vv$ that varies adiabatically as different regions of the twisted homobilayer are explored. From this point of view, we may compute the energy eigenvalues of $H_\alpha(\vv)$ as functions of the adiabatic parameter $\vv$ in the form
\begin{equation}
    E_\alpha^\pm(\vv) = \varepsilon_\alpha^{(0)} + \delta \varepsilon_\alpha(\vv) \pm \left| T_\alpha(\vv) \right|.
\end{equation}
By reinstating $\vv \rightarrow \rr$, the eigenvalues $E_c^-(\vv)$ and $E_v^+(\vv)$ can be interpreted as effective moir\'e potentials acting on the conduction and valence electrons, respectively. These potentials are plotted in Fig.\ \ref{fig:EnergyLand} for a $\theta=2^\circ$ twisted phosphorene bilayer.

The potential $E_c^-(\rr)$ features wells capable of confining conduction electrons, shown in dark blue in the top panel of Fig.\ \ref{fig:EnergyLand}. These potential wells produce either the localized, quantum-dot-like states shown in Fig.\ \ref{fig:WFunctions}, or the quasi-1D states shown in Fig.\ \ref{fig:WFunctions14}, depending on the size of the mSC, as determined by the twist angle. Similarly, $E_v^+(\rr)$ shows potential barriers, depicted in the bottom panel of Fig.\ \ref{fig:EnergyLand} as dark red spots, which can localize holes. Note that in the valence, in addition to the absolute minima that appear at HA regions of the mSC, there are secondary, local minima at AA stacking regions. The appearance of these two competing sets of minima made it necessary to expand the exponential functions describing the $d(\rr)$ dependence of the valence-band parameters up to second order, to correctly capture the relative depths of these potential wells. Note that for both the conduction and valence cases, the potential wells are both anisotropic and irregular, explaining the asymmetric shapes of the charge densities reported in Fig.\ \ref{fig:WFunctions}.

\end{document}